\documentclass[12pt]{article}

  \usepackage{graphicx}
  \usepackage{epsfig}
  \usepackage{amssymb}
  \usepackage{subfigure}
  \usepackage{axodraw}
  \usepackage{amsmath} 

\evensidemargin - .5truecm
\allowdisplaybreaks[1]

\begin{document}

\newcommand{\beq}{\begin{equation}}
\newcommand{\eeq}{\end{equation}}
\newcommand{\beal}{\begin{align}}
\newcommand{\eeal}{\end{align}}
\newcommand{\nn}{\nonumber}
\newcommand{\bea}{\begin{eqnarray}}
\newcommand{\eea}{\end{eqnarray}}
\newcommand{\bfig}{\begin{figure}}
\newcommand{\efig}{\end{figure}}
\newcommand{\bc}{\begin{center}}
\newcommand{\ec}{\end{center}}

\newenvironment{appendletterA}
{
  \typeout{ Starting Appendix \thesection }
  \setcounter{section}{0}
  \setcounter{equation}{0}
  \renewcommand{\theequation}{A\arabic{equation}}
 }{
  \typeout{Appendix done}
 }
\newenvironment{appendletterB}
 {
  \typeout{ Starting Appendix \thesection }
  \setcounter{equation}{0}
  \renewcommand{\theequation}{B\arabic{equation}}
 }{
  \typeout{Appendix done}
 }

%%%%%%%%%%%%%%%%%%%%%%%%%%%%%%%%%%%%%%%%%%%%%%%%%%%%%%%%%%%%%%%%%%%%%%%%
%
%
%\begin{fmffile}{HIGAGA}
%
%
%%%%%%%%%%%%%%%%%%%%%%%%%%%%%%%%%%%%%%%%%%%%%%%%%%%%%%%%%%%%%%%%%%%%%%%%

\begin{titlepage}
\nopagebreak
{\flushright{
        \begin{minipage}{5cm}
         ROME1/1446-06 \\
         IFIC/07-22 \\
        {\tt hep-ph/yymmnnn}
        \end{minipage}        }

}
\renewcommand{\thefootnote}{\fnsymbol{footnote}}
\vskip 2cm
\begin{center}
\boldmath
{\Large\bf The Two Loop Crossed Ladder Vertex Diagram \\[7pt]
with Two Massive Exchanges}\unboldmath
\vskip 1.cm
{\large  U.~Aglietti\footnote{Email: Ugo.Aglietti@roma1.infn.it}},
\vskip .2cm
{\it Dipartimento di Fisica, Universit\`a di Roma ``La Sapienza'' and
INFN, Sezione di Roma, I-00185 Rome, Italy} 
\vskip .2cm
{\large  R. Bonciani\footnote{Email:
Roberto.Bonciani@ific.uv.es}},
\vskip .2cm
{\it Departament de F\'{\i}sica Te\`orica, 
IFIC, CSIC -- Universitat de 
Val\`encia, \\
E-46071 Val\`encia, Spain}
\vskip .2cm
{\large L.~Grassi\footnote{Email: Lorenzo.Grassi@bo.infn.it}},
\vskip .2cm
{\it Dipartimento di Fisica dell'Universit\`a di Bologna, 
and INFN, Sezione di Bologna, \\ 
I-40126 Bologna, Italy} 
\vskip .2cm
{\large E.~Remiddi\footnote{Email: Ettore.Remiddi@bo.infn.it}}
\vskip .2cm
{\it Dipartimento di Fisica dell'Universit\`a di Bologna, 
and INFN, Sezione di Bologna, \\ 
I-40126 Bologna, Italy} 
\end{center}
\vskip 0.7cm

\begin{abstract}
We compute the (three) master integrals for the crossed ladder diagram with 
two exchanged quanta of equal mass. 
The differential equations obeyed by the master integrals are used to generate 
power series expansions centered around all the singular (plus some regular) 
points, which are then matched numerically with high accuracy. 
The expansions allow a fast and precise numerical calculation 
of the three master integrals (better than 15 digits with less than 
30 terms in the whole real axis). 
A conspicuous relation with the equal-mass sunrise in two dimensions 
is found.
Comparison with a previous large momentum expansion is made finding complete 
agreement. 


\end{abstract}
\vfill
\end{titlepage}    

\setcounter{footnote}{0}

\section{Introduction}

Electron-positron linear colliders of next generation with a large energy,  
$E \, = \, {\mathcal O} \left(1\,{\rm TeV}\right)$, 
and a large luminosity, $L \, = \, {\mathcal O} \left(10^{34}{\rm cm}^{-2} {\rm s}^{-1} \right)$, 
will produce an incredible  amount of accurate data \cite{tesla}. 
Theoretical work in the same direction, i.e.
pointing to accurate computations of electroweak cross sections and decay
widths, is therefore mandatory. That means, in practice, to go beyond the
standard one-loop approximation and address the control of the two-loop
quantum corrections.
While in QCD multi-loop computations are often feasible by setting to zero the
small quark masses simplifying then the problem to a massless one, in the
electroweak case setting to zero the masses of the $W$, $Z$, Higgs
or top quark is not possible. One has to face the general problem 
of computing massive multiscale Feynman diagrams. In many cases, the full computation
can be drastically simplified by finding a small expansion parameter
$\eta$, such as the ratio of particle masses and/or kinematical invariants:
\beq
\eta \, = \, \frac{m^2}{M^2} \, \ll \, 1 \, .
\eeq
The heavy algebra entering the higher orders in $\eta$ can be dealt with
using standard algebraic programs such as Form \cite{FORM} 
or Mathematica \cite{mathematica}. In problems
involving many different scales or, equivalently, many different
dimensionless ratios, that is probably the only viable analytical way. 
Another strategy is that of numerical computation of the integrals ``from
the beginning'', such as for example the numerical computation of loop
integrals written in terms of Feynman parameters \cite{Actis:2006xf}. This
approach is not easy to implement when a subset of  particles is massless,
because of troublesome infrared effects.
A different strategy assumes the dominance of the logarithmic terms
over constants and power-suppressed terms in the high-energy cross-sections,
i.e. of terms of the form \cite{Ciafaloni:1998xg}
\beq
\left(\frac{g^2}{16\pi^2}\right)^n \, \log^k \frac{s}{m^2},~~~~~~(k = 1,2,\cdots 2n), 
\eeq
where $g$ is the weak coupling and $m$ is the $W$ or $Z^0$ mass.
The logarithmic terms are computed directly by means of asymptotic expansions
\cite{Denner1l,Pozzorini:2004rm}.
This strategy is similar in spirit to the one used in QCD to construct
for example shower Montecarlo programs \cite{Ciafaloni:2006wb}. 
The logarithmic terms are certainly the dominant ones in the formal limit
$s \to \infty$ (i.e. $s \gg m$), 
but at a fixed energy non logarithmic terms may in general have a significant 
numerical effect.
In cases in which one or (at most) two dimensionless quantities of order one
are involved, an exact analytic computation may be useful.  A model computation
for higher-order electroweak effects is the two-loop form factor in the
degenerate mass limit, 
\beq
m_W \, = \, m_Z \, = \, m_H \, = \, m .
\eeq
In \cite{Aglietti:2003yc, Aglietti:2004tq} two of us have computed the
master integrals of the two-loop electroweak form factor for all diagrams with
the exception of the crossed-ladder with the exchange of two equal mass
quanta --- such as a pair of $W$'s and/or $Z$'s (see fig.~1).  
These master integrals are naturally expressed in the basis of the
harmonic polilogarithms (HPLs), transcendental functions  verifying algebra
properties which largely simplify the transformations \cite{Remiddi:1999ew}.  
Some master integrals involving two or three massive propagators have required 
a generalization of the harmonic polilogarithms basis, involving basic functions 
with square roots --- the so-called generalized harmonic polilogarithms (GHPLs)
\cite{Aglietti:2004tq}.

In this paper we compute the last missing piece in order to obtain an exact analytic
expression of the two-loop electro-weak form factor: the crossed ladder diagram 
with two equal-mass quanta exchanged.
By standard reduction techniques, we find that this topology involves
three master integrals.
At variance with respect to the planar topologies and the non-planar topology 
with at most one massive exchange, it is not possible to represent any of these 
master integrals in terms of harmonic polylogarithms or generalized harmonic 
polylogarithms: elliptic integrals are involved, i.e. integrals  of square roots 
of polynomials of degree three or four \cite{Smirnov}. 
For that reason, we use a different method, already applied in 
\cite{Laporta:2004rb, Pozzorini:2005ff} to evaluate the equal-mass sunrise diagram.
The differential equations obeyed by the master integrals are used
to generate power-series expansions centered around all the singular points of the
differential equations themselves.
It may also be convenient to generate auxiliary expansions around regular points,
which are in principle completely arbitrary.
The series expansions centered around two different points are then matched
numerically in a point belonging to the intersection of the respective domains of
convergence.
In general, we are not able to find a closed analytic expression for the 
master integrals, but only truncated series representations. The latter allow, 
however, a fast and accurate numerical evaluation of the master integrals 
themselves (better than 15 digits with less than 30 terms, 
typically, for the ``accelerated'' versions of the expansions).
The logarithmic terms in the large momentum expansion of one of the master integrals 
(the basic scalar amplitude)  have been obtained in \cite{Smirnov:1998vk}. 
Our expansion is in agreement with these results; we are also able to compute
the power-suppressed corrections.

\section{Power Series Solution of the Differential Equations on the MIs}

In this section we outline the computation of the master integrals
of the two-equal-mass-exchanged crossed-ladder diagram, which is described extensively 
in the next sections.
As anticipated in the introduction, we compute the master integrals
with a general semi-analytical method \cite{Smirnov}, which can be applied 
to arbitrarily complicated cases.
The first steps (sec.~\ref{sec2}) are common to the method used to compute the previous 
electroweak form factor diagrams \cite{Aglietti:2003yc, Aglietti:2004tq}.
By using standard reduction procedures, the problem is shifted to that of 
computing three independent scalar amplitudes, called master integrals (MIs), 
which obey a system of coupled first-order differential equations in 
the evolution variable $x \equiv - s/m^2$, where $s = E_{\rm cm}^2$ is the
Mandelstam variable (sec.~\ref{sec2bis}).
These equations are linear and not homogeneous.
By a suitable choice of the master integral basis, we succeeded in
triangularizing the system to a system of two coupled 
equations (let's say the first two equations on $F_1$ and $F_2$) plus a decoupled equation 
from the previous ones (let's say the third one on $F_3$):
\bea
\label{first_eq}
\frac{dF_1}{dx} &=& A_{11}(x) F_1(x) +  A_{12}(x) F_2(x) + \Omega_1(x);
\\
\label{second_eq}
\frac{dF_2}{dx} &=& A_{21}(x) F_1(x) +  A_{22}(x) F_2(x) + \Omega_2(x);
\\
\label{third_eq}
\frac{dF_3}{dx} &=& A_{31}(x) F_1(x) +  A_{32}(x) F_2(x)+  A_{33}(x) F_3(x) + \Omega_3(x),
\eea
where the $A_{ij}(x)$'s, the coefficients of the associated homogeneous
system, are rational fractions in $x$, while 
the $\Omega_k(x)$'s are known functions (logarithms, polylogarithms etc.). 
We then derive from the system of the two coupled first-order equations a 
single second-order equation in one of the two master integrals involved,
let's say $F_1$ (sec.~\ref{SecOrd}):
\beq
\label{eqsec}
\frac{d^2 F_1}{dx^2} + A (x) \frac{dF_1}{dx} + B(x) F_1(x) = \Omega(x).
\eeq
The resulting equation belongs to the Fuchs class, i.e. it has {\it regular singular} 
points only, including also the point at infinity\footnote{
In general, by a singularity of the differential equation we mean a singularity
in the coefficients $A (x)$ or $B(x)$.
A point $x_0$ is a {\it regular singular point} of a differential equation if its
solutions can be written as singular factor $\sim (x-x_0)^{\alpha}$ or 
$\sim (x-x_0)^{\alpha} \log (x-x_0)$ multiplied by
convergent serieses in a neighborhood of $x_0$.
} \cite{Smirnov}. 
There are {\it four} such singularities, 
all located on the real axis, at:
\beq
x \, = \, 0; ~~~ x \, = \, 8; ~~~ x \, = \, - 1; ~~~ x \, = \, \infty.   
\eeq
The differential equation (\ref{eqsec}) is used to generate power series expansions
centered around all these singular points (sec.~\ref{PowSer}). 
Auxiliary expansions around (arbitrary) regular points may also be of
utility.
Each series is determined by the differential equation up to two
arbitrary coefficients, which must be fixed by some initial or
boundary condition.
A qualitative knowledge of the behaviour of the master integral $F_1(x)$ 
around $x = 0$ is sufficient to fix exactly (in fact analytically) the two 
arbitrary coefficients of the expansion around this point.
The coefficients of the series centered around $x = 0$ (small-momentum expansion) 
are therefore completely determined in analytical way.
The arbitrary coefficients occurring in the serieses centered in the remaining 
points are found by imposing matching conditions in suitable points belonging
to the intersections of the respective domains of convergence. 
Assume that one knows the expansions of $F_1(x)$ around two different points
$x_1$ and $x_2$  
\bea
F_1(x) &=& \sum_{n=0}^{\infty} a_n^{(1)} (x-x_1)^n ~~~ {\rm for} ~~~ 
                                    |x-x_1| \, < \, R_1 \, , \\ 
F_1(x) &=& \sum_{n=0}^{\infty} a_n^{(2)} (x-x_2)^n ~~~ {\rm for} ~~~ 
                                    |x-x_2| \, < \, R_2 \, , 
\eea
with $|x_1 - x_2| \, < \, R_1 + R_2$, so that that the two circles
overlap, and that the coefficients $a_n^{(1)}$ of the first 
series are completely known, while those of the second series are determined 
as (linear) functions of the first two by the underlying differential equation: 
\beq
a_n^{(2)} \, = \, b_n \, a_0^{(2)} \, +  \, c_n \, a_1^{(2)} \, + \, d_n \, .
\eeq
By taking a point $\bar{x}$ in the intersection domain, the matching
conditions give a linear system on the unknowns $a_0^{(2)}$ and $a_1^{(2)}$.
\bea
&& a_0^{(2)} \, \sum_{n=0}^{\infty} b_n \left( \bar{x} - x_2 \right)^n 
\, + \, 
a_1^{(2)} \, \sum_{n=0}^{\infty} c_n \left( \bar{x} - x_2 \right)^n \, =
\nn\\
&& ~~~~~~~~~~~~~ \sum_{n=0}^{\infty} a_n^{(1)} \left( \bar{x} - x_1 \right)^n 
\, - \,  \sum_{n=0}^{\infty} d_n \left( \bar{x} - x_2 \right)^n \, ;
\\
&& a_0^{(2)} \, \sum_{n=1}^{\infty} n \, b_n \left( \bar{x} - x_2 \right)^{n-1} 
\, + \, 
a_1^{(2)} \, \sum_{n=1}^{\infty} n \, c_n \left( \bar{x} - x_2 \right)^{n-1} \, =
\nn\\
&& ~~~~~~~~~~~~~ \sum_{n=1}^{\infty} n \, a_n^{(1)} \left( \bar{x} - x_1 \right)^{n-1}
\, - \,  \sum_{n=1}^{\infty} n \, d_n \left( \bar{x} - x_2 \right)^{n-1} \, .
\eea
Once $a_0^{(2)}$ and $a_1^{(2)}$ are known, one determines the higher-order 
coefficients  $a_n^{(2)}$ for $n > 1$.  
Instead of matching the derivatives of the serieses, one can also take 
a second matching point $\bar{x}'$.

Once the first master integral $F_1$ has been determined, one plugs
its expression into the first-order equation (\ref{first_eq}) of the $2 \times 2$ 
system involving $dF_1/dx$ and determines algebraically $F_2$ (sec.~\ref{SecMI}). 
One finally inserts the expression of the first two MIs $F_1$ and $F_2$
into the third differential equation (\ref{third_eq})
and  determines $F_3$ by quadrature.
Let us remark that one can study all the analytical properties of the MIs 
(analytic continuation, behaviour close to thresholds, asymptotic properties, etc.)
by looking at their power-series expansions.
The latter also offer a suitable and powerful way for the precise and fast 
numerical evaluation of the MIs, for example with a Fortran routine.

In sec.~\ref{relsun} we discuss the relation 
between the differential equation on the master integral $F_1$ for the crossed ladder
and that one for the sunrise
diagram with three equal masses in two space-time dimensions
\cite{Laporta:2004rb}. The ``physical'' origin of this connection is, at
present, unclear.
Finally, in section~\ref{concl} we draw our  conclusions.

\section{Threshold Structure}

The crossed ladder diagram with two equal-mass quanta exchanged has a rather simple 
threshold structure: there is a 2-particle cut on the internal massless lines
for 
\beq
s \, \ge \, 0 
\eeq
and two 3-particle cuts on the internal massless lines and one
of the massive lines for 
\beq
s \, \ge \, m^2.
\eeq
There is also a pseudothreshold for $s \, \le \, 4m^2$.
The complexity of the differential equations and
of the structure of the master integral cannot be
guessed by looking at the threshold structure of the diagram.
That is to be contrasted to the case of the equal-mass sunrise, where
the threshold at $s = 9 m^2$ as well as the pseudothreshold in $s = m^2$
are identifiable as the basic sources of complexity.

\section{Reduction to Master Integrals}
\label{sec2}

\bfig
\vspace*{10mm}
\[\vcenter{
\hbox{\begin{picture}(0,0)(0,0)
\SetScale{1}
\SetWidth{0.5}
\ArrowLine(-80,70)(-60,58)
\ArrowLine(-60,58)(0,23)
\ArrowLine(0,23)(40,0)
\ArrowLine(40,0)(0,-23)
\ArrowLine(0,-23)(-60,-58)
\ArrowLine(-60,-58)(-80,-70)
\DashLine(40,0)(80,0){2}
\SetWidth{2}
\Line(-60,58)(0,-23) 
\Line(-60,-58)(0,23) 
\Text(-32,25)[cb]{{\footnotesize $p_1$}}
\Text(-32,-27)[cb]{{\footnotesize $p_2$}}
\Text(25,3)[cb]{{\footnotesize $Z'$}}
\Text(-15,5)[cb]{{\footnotesize $m$}}
\Text(-15,-5)[cb]{{\footnotesize $m$}}
\end{picture}}
}
\]
\vspace*{18mm}
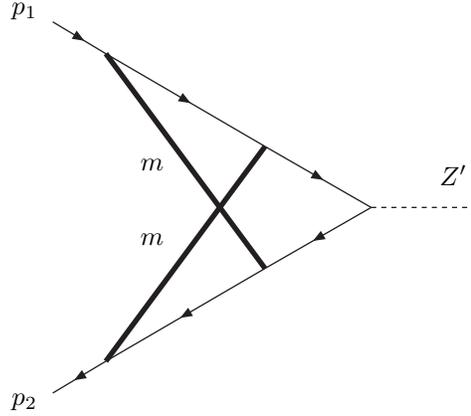
\caption[]{\it Feynman diagram for the annihilation of a pair of massless
fermions with the exchange of two massive quanta with equal mass $m$. The
thin lines represent the massless fermions, while the thick lines represent 
the massive quanta. The outgoing dashed line represents the probe (for
instance a $Z'$).}
\label{fig1}
\efig

By standard decomposition into invariant form factors and rotation of the scalar products, 
one can show that the computation of the two equal-mass crossed ladder 
diagram (see fig.~\ref{fig1}) is equivalent to the computation of the
following independent scalar amplitudes:
\beq
F\left(
n_1,n_2,n_3,n_4,n_5,n_6,s
\right)
\, = \,
\int \frac{S^r}{ P_1^{\,n_1} \,  P_2^{\,n_2} \,  P_3^{\,n_3} \,  P_4^{\,n_4} 
\,  P_5^{\,n_5} \,  P_6^{\,n_6} }
\mathcal{D}^D k_1 \, \mathcal{D}^D k_2,
\eeq
where $D$ is the space-time dimension, the scalar product is defined as
\beq
a \cdot b \, \equiv \,  \vec{a} \cdot \vec{b} \, - \, a_0 \, b_0, 
\eeq
the loop measure is 
\beq
\mathcal{D}^D k \, \equiv \, \frac{1}{\Gamma(3-D/2)} \, \frac{d^D k}{4 \pi^{D/2}} ,
\eeq
with $\Gamma(z)$ the Euler Gamma function. We consider a routing of the
loop momenta $k_1^{\mu}$ and $k_2^{\nu}$ which results in the following denominators:
\bea
P_1 &=& k_1^2 \, + \, m^2 ,
\\
P_2 &=&   k_2^2 \, + \, m^2 ,
\\
P_3 &=&  \left( p_1 \, - \, k_1 \right)^2 ,
\\
P_4 &=&  \left( p_2 \, - \, k_2 \right)^2 ,
\\
P_5 &=&  \left( p_1 \, - \, k_1 \, + \, k_2 \right)^2 ,
\\
P_6 &=&  \left( p_2 \, + \, k_1 \, - \, k_2 \right)^2 ,
\eea
and the following irreducible numerator (scalar product):
\beq
S \, = \, p_2 \cdot k_1 \, .
\eeq 
The indices of the denominators are assumed to be all positive\footnote{
If $n_i \le 0$ for some $i$ we have a sub-topology in which
line $i$ is shrinked to a point.}, $n_i > 0$ 
while the index of the scalar product can be positive  or zero, $r \ge 0$.

The above independent scalar amplitudes, considered as functions of the
integer indices $n_i$'s and $r$, are related to each other by integral
identities obtained in general by means of integration by parts over the loop momenta,
invariance under Lorentz transformations and symmetry relations
coming from the particular mass  distribution of the diagram under
consideration \cite{IBPs}. These identities can be used to express a given
amplitude $F$  as a linear combination of a set of reference amplitudes,
with fixed indices, called master integrals (MI) $F_i$:
\beq
F\left(n_1,n_2,n_3,n_4,n_5,n_6,r\right)
\, = \, 
\sum_{i=1}^N c_i\left(n_1,n_2,n_3,n_4,n_5,n_6,r\right) \, F_i,
\eeq
where the $c_i$'s are known coefficients (rational fractions in $x$ and $D$). 
The above reduction involves the
solution of the integration-by-parts  identities in some recursive way over
the indices. Nowadays the most common algorithm is the ``Laporta algorithm'' \cite{Lap}, 
which we have used for the crossed ladder topology.
That way we have been able to reduce the independent amplitudes 
to a linear combination of three master integrals:
\beq
F\left(n_1,n_2,n_3,n_4,n_5,n_6,s\right)
\, = \, 
\sum_{i=1}^3 c_i\left(n_1,n_2,n_3,n_4,n_5,n_6,s\right) \, F_i
\, + \, \cdots,
\eeq
where the dots denote contributions from master integrals with less
than six denominators, i.e. from subtopologies, which are known
\cite{Aglietti:2003yc, Aglietti:2004tq, fli}.
Let us choose the following basis of master integrals:
\bea
\label{defF1}
F_1(x; \, \epsilon) &=& 
\int \frac{1}{ P_1 \,  P_2 \,  P_3 \,  P_4 \,  P_5 \,  P_6 }
\mathcal{D}^D k_1 \, \mathcal{D}^D k_2 \, ;
\\
F_2(x; \, \epsilon) &=& 
\int \frac{1}{ P_1^2 \,  P_2 \,  P_3 \,  P_4 \,  P_5 \,  P_6 }
\mathcal{D}^D k_1 \, \mathcal{D}^D k_2 \, ;
\\
F_3(x; \, \epsilon) &=& 
\int \frac{S}{ P_1 \,  P_2 \,  P_3 \,  P_4 \,  P_5 \,  P_6 }
\mathcal{D}^D k_1 \, \mathcal{D}^D k_2 .
\eea
The first MI is the basic scalar amplitude, with linear denominators
and no scalar products; the second MI contains the first denominator squared, 
while the third MI involves an irreducible numerator.
By direct inspection, one finds that the above MI's do not contain any pole in
$D-4$ because they are ultraviolet finite and, since the radiated quanta are 
massive, there is an effective infrared cut-off $\approx m$ which renders the MIs 
infrared finite as well.
Let us therefore set $D \, = \, 4$ from now on and omit this argument. 

\section{The System of Differential Equations}
\label{sec2bis}

Once the reduction to master integrals has been achieved,
the following step is the actual computation of the master
integrals themselves.
We use the differential equation method \cite{DiffEq}, which involves
taking a derivative with respect to
\beq
x \, = \, - \, \frac{s}{m^2} ,
\label{xdef}
\eeq
at fixed $m^2$, of the master integral $F_i$:
\beq
x \, \frac{\partial F_i}{\partial x}  \, 
\, = \, \frac{1}{2} \, p_1^{\mu} \, \frac{\partial  F_i }{\partial p_1^{\mu} }   
\, = \, \frac{1}{2} \, p_2^{\mu} \, \frac{\partial  F_i }{\partial p_2^{\mu} } \, .
\label{deq}
\eeq
$q = p_1 + p_2 $ the momentum of the probe and $q^2=-s$, $s$ being the squared c.m. energy.
By taking the derivative inside the integral in eq.~(\ref{deq}), 
various amplitudes are generated, which are reduced again
to master integrals as discussed in the previous section.
We then obtain in general a system of linear differential 
equations with variable coefficients, of the form:
\beq
\label{eqs}
\frac{d}{d x} F_i(x)
\, = \, \sum_{j=1}^3 f_{ij}(x) \, F_j(x) \, + \, N_i(x) 
~~~~~~~~~~~~~ (i \, = \, 1,2,3) \, .
\eeq
By changing the basis for the master integrals, the functions  $f_{ij}(x)$
and $N_i(x)$ are transformed into new functions.
A general basis for the MI's involves
a system of three coupled differential equations,
which are equivalent to a single, third-order
differential equation, whose solutions, as well 
known, are rather difficult to find.
With the basis given in the previous section, the system comes
out to be triangular with a $2 \times 2$ block.
The coefficients of the associated homogeneous system read, in this basis:
\begin{align}
\hspace*{-4mm}
f_{11}(x) & = - \frac{2}{x} ; 
& f_{12}(x) & = \frac{2}{x} ;
& f_{13}(x) & = 0 ;
\\
\hspace*{-4mm}
f_{21}(x) &= - \frac{1}{2x} + \frac{1}{3(x+1)} + \frac{1}{6(x-8)} ;
& f_{22}(x) &= - \frac{1}{x+1} - \frac{1}{x-8} ;
& f_{23}(x) &= 0 ;
\\
\hspace*{-4mm}
  f_{31}(x) &= - \frac{1}{6x} ;
& f_{32}(x) &= \frac{1}{3} \left( 1 + \frac{1}{x} \right) ;
& f_{33}(x) &= - \frac{1}{x} ;
\end{align}
while the known terms, related to the subtopologies, are:
\bea
N_1(x) &=& 0 ;
\\
N_2(x) &=& \frac{1}{16 x}
\Biggl[
\frac{1}{2}  H(0,0; x)
 -  \frac{9}{4}  H(0,-1; x)
 +  \frac{1}{\sqrt{x(4-x)}}  H(r,0; x)
 +  \frac{\pi^2}{4}
\Biggr] 
\nonumber \\
&  &
- \frac{1}{4(x+1)}
\Biggl[
    \frac{1}{9}  H(0,0;  x) 
 -  \frac{1}{2}  H(0,-1;  x)
 +  \frac{5}{12 \sqrt{x(4-x)}}  H(r,0;  x) 
 + \frac{\pi^2}{18}
\Biggr]  
\nonumber \\
& &
- \frac{1}{8(x-8)}
\Biggl[
    \frac{1}{36}  H(0,0;  x) \!
 - \!  \frac{1}{8}  H(0,-1;  x) \!
 - \!  \frac{1}{3\sqrt{x(4-x)}}  H(r,0;  x) \!
 +  \! \frac{\pi^2}{72}
\Biggr] ;
\label{N2} \\
N_3(x) &=& 
- \frac{1}{48x^2} 
\Biggl[
\frac{1}{2}  H(-1,0,0;  x)
 +  H(r,r,0;  x)
 +  \frac{\pi^2}{4}  H(-1;  x)
\Biggr] \, .
\label{N3} 
\eea
The known terms are written in terms of generalized harmonic
polylogarithms \cite{Aglietti:2004tq}.
Since $f_{13}(x) = 0$ and $f_{23}(x) = 0$,
$F_3$ decouples from the system of the first two MI's $F_1$ and $F_2$. 
It is therefore natural to compute $F_1$ and $F_2$ first.
Once the latter are known, the third master integral $F_3$ can be computed by 
quadrature --- i.e. by integration of the first-order differential equation
(\ref{eqs}) with $i \, = \, 3$.

\subsection{Master Integrals Close to Zero External Momentum}

Since $p_1^2 \, = \, p_2^2 \, = \, 0$, 
\beq
q^2 \, = \, 2 \, p_1 \cdot p_2
\eeq
and the limit 
\beq
\label{limitxto0}
x \, \to \, 0
\eeq
is equivalent to the limit $p_1 \cdot p_2 \, \to \, 0$. 
However, limit (\ref{limitxto0}) also implies the {\it stronger} limits 
$p_1^{\mu} \to 0$ and $p_2^{\nu} \to 0$, which reduce the vertex topology
to a vacuum one.
That can be seen for example by introducing Feynman parameters $x_1, \, x_2 \cdots x_5$ 
for the vertex amplitude and integrating over the loop momenta $k_1^{\mu}$ and $k_2^{\nu}$.
After that, the integrand can depend on relativistic invariants only, i.e. just on $q^2$.
One then performs analytic continuation to Euclidean space, where the limit $q^2 \to 0$
implies $q^{\mu} \to 0$, i.e. $p_2^{\mu} = - \, p_1^{\mu}$.
The diagram can then depend only on $p_1^2 = 0$ and therefore must be equal to that
one computed for $p_1^{\mu} = 0$, i.e. to the corresponding vacuum amplitude.
In particular, the MI's reduce to the following vacuum amplitudes:
\bea
F_1(x = 0) & = &
\int \frac{1}{ (k_1^2 \, + \, m^2) \, (k_2^2 \, + \, m^2)  \,  k_1^2 \,  k_2^2 \,  \left[ (k_1-k_2)^2 \right]^2 }
\, \mathcal{D}^4 k_1 \, \mathcal{D}^4 k_2 ;
\\
F_2(x = 0) & = &
\int \frac{1}{ (k_1^2 \, + \, m^2)^2 \, (k_2^2 \, + \, m^2)  \,  k_1^2 \,  k_2^2 \,  \left[ (k_1-k_2)^2 \right]^2  }
\, \mathcal{D}^4 k_1 \, \mathcal{D}^4 k_2 ;
\\
F_3(x = 0) & = & 0 .
\eea
The above integrals can be computed exactly by using the identities
\beq
\frac{1}{k_i^2} \, \frac{1}{k_i^2+m^2} \, = \, \frac{1}{m^2 \, k_i^2} - \frac{1}{ m^2 \, (k_i^2+m^2) }.
\eeq
For our purposes, that is actually not necessary.
By shrinking the massive lines in the infrared regions
$k_1^2,~k_2^2 \ll m^2$, we obtain:
\bea
F_1(x = 0) & \approx & \frac{1}{m^4} \,
\int^{\Lambda_{UV}} \frac{1}{ k_1^2 \,  k_2^2 \,  \left[ (k_1-k_2)^2 \right]^2 }
\mathcal{D}^4 k_1 \, \mathcal{D}^4 k_2 ;
\\
F_2(x = 0) & \approx & \frac{1}{m^6} \,
\int^{\Lambda_{UV}} \frac{1}{ k_1^2 \,  k_2^2 \,  \left[ (k_1-k_2)^2 \right]^2  }
\mathcal{D}^4 k_1 \, \mathcal{D}^4 k_2,
\eea
where an ultraviolet cutoff $\Lambda_{UV} \approx m$ is assumed.
The integrands above are invariant under the limit $\lambda \, \to \, 0$ in $D = 4$,
where $\lambda$ is introduced through the rescaling 
\bea
k_1^{\mu} &\to& \lambda \, k_1^{\mu} ,
\\
k_2^{\mu} &\to& \lambda \, k_2^{\mu}.
\eea
All that implies an over-all logarithmic soft divergence\footnote{
There is not any soft sub-divergence, as can be seen by power-counting
under the rescalings $k_1^{\mu} \to \lambda \, k_1^{\mu} , ~~ k_2^{\mu} \to k_2^{\mu}$
and $k_1^{\mu} \to k_1^{\mu} , ~~ k_2^{\mu} \to \lambda \, k_2^{\mu}$.
}.
There is also a logarithmic collinear singularity 
for $k_1^{\mu} \propto k_2^{\mu}$ related to the denominator $\left[ (k_1-k_2)^2 \right]^2$.
As a consequence\footnote{
If one sets $x = 0$ from the very beginning, $1/(D-4)$ poles
of infrared nature do appear in $F_1$ and $F_2$.
}:
\beq
F_{1,2}(x) \, \sim \, \log^2 x \, , ~~~~~
F_{3}(x) \, \sim \, 0 ~~~~~
{\rm for}~ x \to 0.
\eeq
That implies in particular that no pole terms $\approx 1/x$ can appear.
As we are going to show explicitly, this qualitative information
is sufficient for the solution of the system at small external momenta.
Finally, let us note that we can set $m = 1$
from now on without loosing any information ($m$ is an over-all scale).

\section{Second-Order Differential Equation for $F_1(x)$}
\label{SecOrd}

It is convenient to transform the system of differential equations
for $F_1$ and $F_2$ into a single second-order equation for
$F_1$ only. 
The procedure is standard: one takes an additional derivative
with respect to $x$ on both sides of eq.~(\ref{eqs}) with $i \, = \, 1$;
$d F_1 /d x$ and $d F_2 /d x$ are replaced by the r.h.s. of eqs.~(\ref{eqs}) 
with $i \, = \, 1$ and $i \, = \, 2$ respectively; finally $F_2$
is replaced by its expression in terms of $d F_1 /d x$ and $F_1$, coming 
from the first of eqs.~(\ref{eqs}).
We obtain:
\beq
\label{lei}
\frac{ d^2  F_1 }{ d x^2 } (x) 
 +  A(x)  \frac{ d F_1 }{ d x } (x) 
 +  B(x)  F_1(x)
 +  C(x)  =  0,
\eeq
where:
\bea
A(x) &=&
    \frac{3}{x}  
 +  \frac{1}{x+1} 
 +  \frac{1}{x-8} ; 
\\
B(x) &=& 
    \frac{1}{x^2} 
 +  \frac{9}{8x} 
 -  \frac{4}{3(x+1)} 
 +  \frac{5}{24(x-8)} ; 
\\
C(x) &= &
- \left[
    \frac{1}{16x^2}
 -  \frac{7}{128x}
 +  \frac{1}{18(x+1)}
 -  \frac{1}{1152(x-8)}
\right] H(0,0; x) 
\nn\\
& &
+ \left[
    \frac{9}{32x^2}
 -  \frac{63}{256 x}
 +  \frac{1}{4(x+1)}
 -  \frac{1}{256(x-8)}
\right] H(0,-1;  x) 
\nn\\
& &
- \left[
    \frac{1}{8x^2}
 -  \frac{7}{32x}
 +  \frac{5}{24(x+1)}
 +  \frac{1}{96(x-8)}
\right] \frac{1}{\sqrt{x(4-x)}}  H(r,0;  x)
\nn\\
& &
- \frac{\pi^2}{4} 
\left[
    \frac{1}{8x^2}
 -  \frac{7}{64x}
 +  \frac{1}{9(x+1)}
 -  \frac{1}{576(x-8)}
\right].
\label{eq54}
\eea
Let us make a few remarks:
\begin{enumerate}
\item
the coefficient $A(x)$ of $F_1'(x)$ contains simple poles, 
while the coefficient $B(x)$ of $F_1(x)$ contains at most double poles.
That is the necessary and sufficient condition on the differential equation 
to have regular singular points only \cite{Smirnov};
\item
the denominators entering the coefficients are:
\beq
\frac{1}{x};~~~~\frac{1}{x+1};~~~~\frac{1}{4-x};~~~~\frac{1}{x-8}.
\eeq
The last denominator was not expected {\it a priori}, as it does not correspond to any 
threshold or pseudothreshold of the diagram.
The denominator $1/(4-x)$ only appears in the known function $C(x)$ and it is related to 
the sub-topologies with the pseudothreshold in $s \, = \, - \, 4 m^2$;
\item
the known term $C(x)$ contains GHPL's (see appendix \ref{appa}) of weight
at most  two;
\item
the GHPL $H(r,0; \, x)$, containing a single square-root basic function,
has a coefficient involving a square root $1/\sqrt{x(4-x)}$ to
ensure reality of the solution across the pseudothreshold located 
at $x = 4$. 
\end{enumerate}

\section{Power-Series Solution of the Second-Order Differential Equation for $F_1(x)$}
\label{PowSer}

The second-order differential equation (\ref{lei}) for $F_1$ 
is by far too complicated to be solved in a closed analytical form. 
We therefore look for solutions in the form of power series expansions 
centered around some points on the real axis. Let us recall that it 
is in any case necessary to consider the expansions around all the singular 
points, as the latter are by definition outside the convergence region 
of the ordinary expansions at regular points. Thanks to the differential 
equation satisfied by the considered functions, expansions around 
singular or regular points can be performed almost in the same way. 

Since the coefficients $A$ and $B$ vanish for $x \, = \, 0, \, - 1, \, 8$, we 
conclude that these points are singular points for the differential equation.
Also $x = \infty$ is a singular point for the differential equation,
as can be seen by changing variable to $y \, = \, 1/x$ and studying the limit 
$y \, \to \, 0$. 

\subsection{Expansion Around $x \, = \, 0$ --- Small Momentum Expansion}

In this section we consider the expansion of the master integral $F_1(x)$
around $x \, = \, 0$.
Since the nearest singularity to the origin is located
in $x = - 1$, we expect a radius of convergence 
\beq
R_0 \, = \, 1,
\eeq
i.e. convergence in the complex $x$-plane for
\beq
\label{circle}
|x| \, < \, 1.
\eeq
For real $x$, that means:
\beq
- \, 1 \, < \, x \, < \, 1.
\eeq
Because of linearity, the general solution of the complete inhomogeneous
equation can be written as the superposition of the
{\it general} solution of the associated {\it homogeneous} equation
---  i.e. eq.~(\ref{lei}) with $C(x) = 0$, 
\beq
\label{lei_omog}
\frac{ d^2  F_1^{(0)} }{ d x^2 }(x) 
\, + \, 
A(x) \, \frac{ d F_1^{(0)} }{ d x }(x) 
\, + \, 
B(x) \, F_1^{(0)}(x)
= \, 0,
\eeq
plus a {\it particular} solution of the {\it complete} equation:
\beq
\label{usatasempre}
F_1(x) \, = \,  F_1^{(0)}(x) \, + \, \bar{F}_1(x).
\eeq
We will use systematically eq.~(\ref{usatasempre}) in the following expansions.

\subsubsection{Homogeneous Equation}

Let us first consider the associated homogeneous equation.
Since zero is a ``singular regular point'', we can look for
solutions having the form of a singular function in $x = 0$,
$S(x)$, multiplied by a regular (i.e. convergent) power-series\footnote{
This case is to be contrasted to that of an ``irregular singular point'',
in which the singularity of the differential equation is so strong
that no factorization of the singularity of the form (\ref{singfac})
is possible.
In the latter case, one typically obtains series with an infinite number
of negative powers or asymptotic (divergent) expansions.
} :
\beq
\label{singfac}
F_1^{(0)}(x) \, = \, S(x) \, \sum_{n=0}^{\infty} A_n \, x^n,
\eeq
where $A_n$ are numerical coefficients determined from the
differential equation itself and from some initial or
boundary conditions.
The function $S(x)$, giving the leading singularity for $x \to 0$,
is assumed to be of power-like form
\beq
\label{forma}
S(x) \, = \, x^{\alpha},
\eeq
and solves the limit of the homogeneous equation for $x \to 0$:
\beq
S''(x)  +  \frac{3}{x}  S'(x)  +  \frac{1}{x^2}  S(x) \, = \, 0.
\eeq
By inserting a solution of the form (\ref{forma}) we obtain the indicial
equation $(\alpha + 1)^2 = 0$ with a double zero in $ \alpha = - 1$, 
implying two independent pre-factors of the form:
\beq
S(x) \, = \, \frac{1}{x}, ~~ \frac{\log x}{x} \, .
\eeq
The most general solution of the homogeneous differential equation 
is therefore of the form:
\beq
\label{gen_omog}
F_1^{(0)}(x)  =  \sum_{n=-1}^{\infty} a_n  x^n 
 +  \log x  \sum_{n=-1}^{\infty} b_n  x^n \, , 
\eeq
where we have absorbed a $1/x$ factor inside the series by
defining:
\beq
a_n  \equiv  A_{n+1}.
\eeq
By expanding the differential equation (\ref{lei_omog}) around $x = 0$
and substituting the series representation in eq.~(\ref{gen_omog}), 
we can obtain recursively all the desired coefficients. 
The first few are: 
\begin{align}
a_0 & =  -  \frac{1}{4} a_{-1}  -  \frac{3}{8} b_{-1} \, ;
& b_0 & = -  \frac{1}{4} b_{-1} \, ;
\\
a_1 & =  \frac{5}{32} a_{-1}  +  \frac{33}{128} b_{-1} \, ;
& b_1 & = \frac{5}{32} b_{-1} \, ;
\\
a_2 & =  -  \frac{7}{64} a_{-1}  -  \frac{25}{128} b_{-1} \, ;
& b_2 & = -  \frac{7}{64} b_{-1} \, ;
\\
a_3 & =  \frac{173}{2048} a_{-1}  +  \frac{2561}{16384} b_{-1} \, ;
& b_3 & = \frac{173}{2048} b_{-1} \, .
\end{align}
Let us make a few remarks:
\begin{itemize}
\item
setting to zero for example $a_{-1}$ and $b_{-1}$, all the higher-order
coefficient vanish and we obtain the trivial solution: that is the 
homogeneity property; 
\item
there is a triangular structure: all the coefficients $b_i$ are proportional to 
the lowest-order one $b_{-1}$, implying that by setting $b_{-1} = 0$
we obtain a solution without the series with the logarithmic prefactor
and with the simple pole in $x = 0$ only.
The coefficients $a_i$ depend instead on both $a_{-1}$ and $b_{-1}$,
implying that by setting $a_{-1} = 0$ one obtain a solution
containing both serieses;
\item
A first independent solution can be obtained by taking for example
$a_{-1} = 1$ and $b_{-1} = 0$, resulting in a function without
logarithmic terms.
A second independent solution can be obtained by taking 
$a_{-1} = 0$ and $b_{-1} = 1$, resulting in a function
having both the power and the logarithmic terms.
With a pictorial language, we may say that the logarithmic series
``feeds'' the standard one, while the vice-versa is not true.
\item
The singularity of the differential equation at $x \, = \, 0$
produces a logarithmic branch point in the solution at $x \, = \, 0$
whenever $b_{-1} \ne 0$.
\end{itemize}

\subsubsection{Complete Equation}

Let us now consider the complete equation (\ref{lei}).
We have to expand around $x \, = \, 0$ 
also the inhomogeneous term $C(x)$, which is known 
(see appendix \ref{appb}):
\beq
\label{known0}
C(x)  \, = \, 
\frac{1}{x^2}
\left[
\sum_{n=0}^{\infty} k_n  x^n
 +  \log x \sum_{n=0}^{\infty} q_n  x^n
 +  \log^2 x \sum_{n=0}^{\infty} r_n  x^n \, .
\right]
\eeq
The known term has a double pole in $x = 0$,
multiplied also by a single or a double logarithm of $x$.
The expected radius of convergence of the multiplying series is one: 
\beq
R \,  = \, 1.
\eeq
The explicit expressions of the lowest-order coefficients
read:
\begin{align} 
k_{0} & =  \frac{1}{8}  -  \frac{\pi^2}{32} \, ;
&
q_{0} & = -  \frac{1}{16} \, ;
&
r_{0} & = -  \frac{1}{32} \, ;
\\
k_{1} & =  \frac{23}{288}  +  \frac{7}{256}  \pi^2 \, ;
&
q_{1} & = \frac{19}{192} \, ;
&
r_{1} & = \frac{7}{256} \, ;
\\
k_2 & = -  \frac{3931}{28800}  -  \frac{57}{2048} \pi^2 \, ;
&
q_2 & = -  \frac{671}{7680} \, ;
&
r_2 & = -  \frac{57}{2048} \, ;
\\
k_3 & = \frac{1789247}{11289600}  +  \frac{455}{16384} \pi^2 \, ;
&
q_3 & = \frac{38791}{430080} \, ;
&
r_3 & = \frac{455}{16384} \, .
\end{align}
Let us look for a particular solution of (\ref{lei}) of the form:
\beq
\bar{F}_1(x)  =  
\sum_{n=-1}^{\infty} p_n  x^n
 +  \log x  \sum_{n=-1}^{\infty} q_n  x^n
 +  \log^2 x  \sum_{n=0}^{\infty} c_n  x^n \, .
\eeq
By inserting the above form of the solution into the
differential equation expanded around $x \, = \, 0$ with the known
term $C(x)$ given by the series expansion in eq.~(\ref{known0}), 
we obtain for the coefficients:
\bea
p_0 &=& -  \frac{1}{4}  p_{-1}  - 
\frac{3}{8}   q_{-1}  +  \frac{\pi ^2}{32}  -  \frac{1}{16} \, ;
\\
q_0 &=& - 
\frac{1}{4}  q_{-1}  -  \frac{1}{16} \, ;
\\
c_0 &=&
\frac{1}{32} \, ;
\\
&~& 
\nn
\\
p_1 &=&
\frac{5}{32} p_{-1}  +  \frac{33}{128} q_{-1}  -  \frac{\pi ^2}{64} 
 +  \frac{5}{576} \, ;
\\
q_1 &=&
\frac{5}{32}   q_{-1}  +  \frac{1}{96} \, ;
\\
c_1 &=& - 
\frac{1}{64} \, ;
\\
&~& 
\nn
\\
p_2 &=& - 
\frac{7}{64}  p_{-1}  -  \frac{25}{128}  q_{-1}  +  \frac{13}{1152}  \pi ^2
 -  \frac{1093}{518400} \, ;
\\
q_2 &=& - 
\frac{7}{64}  q_{-1}  -  \frac{121}{17280} \, ;
\\
c_2 &=&
\frac{13}{1152} \, ;
\\
&~& 
\nn
\\
p_3 &=&
\frac{173}{2048} p_{-1}  +  \frac{2561}{16384} q_{-1}  -  \frac{5 \pi ^2}{576} 
 + 
\frac{649}{635040} \, ;
\\
q_3 &=&
\frac{173}{2048}  q_{-1}  +  \frac{95}{24192} \, ;
\\
c_3 &=& - 
\frac{5}{576} \, ;
\\
&~& 
\nn
\\
p_4 &=& - 
\frac{563}{8192}   p_{-1}  -  \frac{42631}{327680}  q_{-1} 
 +  \frac{407}{57600}  \pi ^2  -  \frac{3217}{6720000} \, ;
\\
q_4 &=& - 
\frac{563}{8192}  q_{-1}  -  \frac{1141}{432000} \, ;
\\
c_4 &=& \frac{407}{57600} \, .
\eea
The coefficients $c_i$ of the double-logarithmic terms
are completely determined, while the remaining ones
$p_i$ and $q_i$ are fixed once two arbitrary
coefficients, such as for example $p_{-1}$ and $q_{-1}$, have been 
fixed. That is exactly the same arbitrariness that we have already 
found for the associated homogeneous equation.
Since the general solution of the latter has already been found, 
we need only to find a particular,
i.e. a single solution of the complete equation.
Let us choose for example:
\beq
p_{-1}  =  0~~~{\rm and}~~~ q_{-1}  =  0 \, .
\eeq
With the following {\it arbitrary} choice, the particular solution does
not contain any pole term and its numerical coefficients are completely fixed:
\begin{align}
\label{inizio}
p_0 & = -  \frac{1}{16}  + 
\frac{\pi ^2}{32} \, ;
&
q_0 &= - 
\frac{1}{16} \, ;
&
c_0 &=
\frac{1}{32} \, ;
\\
p_1 &= 
\frac{5}{576}  -  \frac{\pi ^2}{64} \, ,
&
q_1 &= 
\frac{1}{96} \, ;
&
c_1 &= - 
\frac{1}{64} \, ,
\\
p_2 &= - 
\frac{1093}{518400}  +  \frac{13 \pi ^2}{1152} \, ;
&
q_2 &= - 
\frac{121}{17280} \, ,
&
c_2 &=
\frac{13}{1152} \, ;
\\
p_3 &=
\frac{649}{635040}  -  \frac{5 \pi ^2}{576} \, ,
&
q_3 &=
\frac{95}{24192} \, ;
&
c_3 &= - 
\frac{5}{576} \, ,
\\
p_4 &= - 
\frac{3217}{6720000}  +  \frac{407 \pi ^2}{57600} \, ;
&
q_4 &= - 
\frac{1141}{432000} \, ,
&
c_4 &=
\frac{407}{57600} \, .
\label{fine}
\end{align}
The general solution of the complete equation is therfore:
\beq
F_1(x)
=
\sum_{n=-1}^{\infty} a_n  x^n
 + \log x   \sum_{n=-1}^{\infty} b_n  x^n
 + \sum_{n=0}^{\infty} p_n  x^n
 + \log x   \sum_{n=0}^{\infty} q_n  x^n
 + \log^2 x  \sum_{n=0}^{\infty} c_n  x^n \, .
\eeq
To uniquely determine the solution, one has to impose
some boundary or initial conditions.
As already discussed in the previous section, $F_1(x)$
can have at most a logarithmic singularity for $x \to 0$,
implying that the coefficients of the power singularities
must vanish:
\beq
a_{-1}  =  0 , ~~~ b_{-1}  =  0 \, ;
\eeq
and the solution of the homogeneous equation to be selected, reduces 
to the trivial one.
The particular solution $\bar{F}_1$ of the complete equation that we have 
chosen is therefore the expansion of the master integral $F_1$
as defined by eq.~(\ref{defF1}):
\beq
\label{Eq0}
F_1(x)
= \sum_{n=0}^{\infty} p_n  x^n
 + \log x   \sum_{n=0}^{\infty} q_n  x^n
 + \log^2 x  \sum_{n=0}^{\infty} c_n  x^n,
\eeq
where the coefficients are given in eqs.~(\ref{inizio}-\ref{fine})

Let us stress that we have been able to obtain the complete analytical
expression of the coefficients of this small-momentum expansion
because of the knowledge of the MI at small momentum transferred.
As we are going to show in the next sections, the absence of a similar
knowledge in other expansion points will limit us to a numerical
estimate of the corresponding coefficients. 

As expected from the threshold structure,
the solution given in eq.~(\ref{Eq0}) is real for $x \, > \, 0$ (space-like region) 
and has a non-vanishing imaginary part for $x \, < \, 0$ (time-like region). 
The latter can be determined by simply giving the prescription for the $\log x$ factor. 
Since $s \to s + i\epsilon$ with $\epsilon \, = \, + 0$, eq.~(\ref{xdef}) gives for $x \, < \, 0$:
\beq
x \to - | x | - i \epsilon \, .
\eeq
Consequently, we have:
\bea
\log x   & \to & \log | x | - i \pi \, ; 
\nn 
\\
\label{prime2}
\log^2 x & \to & \log^2 | x | - \pi^2 - 2 i \pi \log | x | \, .
\eea
Therefore $F_1(x)$ becomes complex for $x \, < \, 0$ with:
\bea
{\rm Re} \, F_1(x) & = & \sum_{n=0}^{\infty} p_n  x^n 
+ \log |x| \sum_{n=0}^{\infty} q_n  x^n + 
\left( \log^2 |x| - \pi^2 \right) \sum_{n=0}^{\infty} c_n  x^n \, ;
\\
{\rm Im} \, F_1(x) & = & - \pi \, 
\left( 
\sum_{n=0}^{\infty} q_n \, x^n
+ 2 \log | x | \, \sum_{n=0}^{\infty} c_n \, x^n 
\right) \, .
\eea
In the following sections, we will ``move'' on the real axis constructing the solution
from $x \, = \, 0$ to $x \, = \, \infty$; then with an analytic continuation we will switch 
to $x \, = \, - \, \infty$ and finally move back to $x \,= \, 0$ through negative values 
of $x$.

\subsubsection{Improved Expansion --- Bernoulli Variable}

A series for $F_1$ with better convergence properties than the previous one in $x$ can 
be constructed by expanding in the Bernoulli variable \cite{bernoulli}
\bea
\label{bernoulli}
t &\equiv& \log \left( 8 \, \frac{1 \, + \, x}{8 \, - \, x} \right)
\, = \, 
\frac{9}{8} x - \frac{63}{128} x^2 + \frac{171}{512} x^3
- \frac{4095}{16384} x^4 + \frac{32769}{163840} x^5 + {\mathcal O}\left(x^6\right)
\\
&=& 1.125 x - 0.492187 x^2 + 0.333984 x^3 - 0.249938 x4 
      + 0.200006 x^5 + {\mathcal O}\left(x^6\right).
\nonumber
\eea
Eq.~(\ref{bernoulli}) realizes a one-to-one map between $x \in (-1,8)$ and 
$t \in (-\infty,\infty)$.
Furthermore,
\beq
t \, = \, 0 \, \iff \, x \, = \, 0, 
\eeq
while 
\bea
t &\to& - \, \infty ~~~~ {\rm for} ~~~~ x \, \to - \, 1^+
\nn\\ 
t &\to& + \, \infty ~~~~ {\rm for} ~~~~ x \, \to + \, 8^- .
\eea
$t$ diverges logarithmically when $x$ approaches the singularities
of the differential equation closest to the origin.
Since this variable ``follows'' the singularities of the
differential equation, we expect a faster convergence with the
order of truncation in $t$ rather than in $x$.
The inverse of eq.~(\ref{bernoulli}) reads:
\beq
\label{xdit}
x  \, = \, \frac{8 ( e^t - 1) }{e^t + 8} \, = \, \sum_{n=1}^{\infty} c_n \, t^n \, .
\eeq
The first few terms are:
\bea
x &=& \frac{8}{9}  t
 +  \frac{28 }{81}  t^2
 +  \frac{44 }{729}  t^3
 -  \frac{35 }{6561}  t^4
 -  \frac{1733 }{295245}  t^5
 +  O\left( t^6 \right)
\\
&=&
0.888888 t \, + \, 0.345679 \, t^2 \, + \, 0.0603566 \, t^3 
\, - \, 0.00533455 \, t^4 \, - \, 0.00586970 \, t^5 \, + \, O\left( t^6 \right).
\nn\eea 
The radius of convergence of the above series is determined by looking at the
singularities of $x \, = \, x(t)$ in the complex $t$-plane, located in
\beq
t_k \,  = \, 3 \log 2 + i(2k+1) \pi,
\eeq
where $k$ is an integer.
The closest singularities to the origin, for $k \, = \, 0, - 1$, give
\beq
\label{Rt}
r_0 \, = \, \sqrt{\pi^2 + 9 \log^2 2} \, \approx \, 3.76745 .
\eeq
We now substitute the r.h.s. of eq.~(\ref{xdit})
in the series expansion for $F_1(x)$ obtained previously 
and finally expand in powers of $t$, to have\footnote{
Let us note that we did not express the logarithmic pre-factors $\log x$ and 
$\log^2 x$ as series in $t$, because there was no practical advantage for 
doing that. 
}:
\beq
\label{F1t}
\tilde{F}_1(t) \, \equiv \, F_1(x(t)) \, = \, 
\sum_{n=0}^{\infty} \alpha_n \, t^n
\, + \,
\log x  \, \sum_{n=0}^{\infty} \beta_n \, t^n
\, + \,
\log^2 x \, \sum_{n=0}^{\infty} \gamma_n \, t^n,
\eeq
where the coefficients $\alpha_n$, $\beta_n$ and $\gamma_n$
are determined from $a_n$, $b_n$ and $c_n$ respectively.
The first few orders read:
\begin{align}
\alpha_0 &= - \, \frac{1}{16} \, + \, \frac{\pi^2}{32} \, ;
&
\beta_0 \, &= \,  - \, \frac{1}{16} \, ;
&
\gamma_0 \, &= \,  \frac{1}{32} \, ,
\\
\alpha_1 &= \, \frac{5}{648} \, - \, \frac{\pi^2}{72} \, ;
&
\beta_1 \, &= \,  \frac{1}{108} \, ,
&
\gamma_1 \, &= \,  - \, \frac{1}{72} \, ;
\\
\alpha_2 &= \, \frac{3503}{2624400} \, + \, \frac{41}{11664} \, \pi^2 \, ,
&
\beta_2 \, &= \,  - \, \frac{169}{87480} \, ;
&
\gamma_2 \, &= \, \frac{41}{11664} \, .
\end{align}
We have computed $\tilde{F}_1(t)$ as a function of $t$ (eq.~(\ref{F1t})) by 
substituting the series (\ref{xdit}) into the series (\ref{Eq0}).
The problem now is that of computing the radius of convergence 
$\rho_0$ of the series expansion for $\tilde{F}_1(t)$.
The ``internal'' series for $x \, = \, x(t)$ converges inside the circle in 
the $t$-plane of radius $r_0$ given in eq.~(\ref{Rt});
the ``external'' series for $F(x)$ converges in the unitary circle $|x| \, < \, 1$ 
in the $x$-plane (eq.~(\ref{circle})).
As well known, power series always converge inside circles.
The point is that a circle in the $t$-plane in general is {\it not} mapped by 
the function $x \, = \, x(t)$ into a circle in the $x$-plane.
Our problem has the following geometrical formulation:
one has to find the largest circle in the $t$-plane satisfying $|t| < r_0$,
which is mapped inside the circle of unitary radius in $x$-plane by the transformation 
$x \, = \, x(t)$ \cite{markusevic}.
On the unitary circle $x = \exp i \varphi$ and one has to look for a minimum
over $\varphi$ of
\beq
|t| \, = \, \left| \log \left( 8 \frac{1 + \exp i \varphi }{8 - \exp i \varphi} \right) \right|
\, \ge \, \log \frac{16}{7} \, < \, r_0,
\eeq
where the minimum is obtained for $\varphi = 0$.
We then find\footnote{
We have shown that the convergence radius of the $t$-series of $F_1$ is {\it not smaller} than $\rho_0$,
but it can actually be larger, depending on possible elimination of singularities.
That can be illustrated with the following (rather trivial) example.
Let us consider the expansion around $x \, = \, 0$ of a differential equation
having a solution of the form
\beq
\phi(x) \, = \, \log \frac{8(1+x)}{8-x}.
\eeq
If we go to the Bernoulli variable $t$ defined in eq.~(\ref{bernoulli}),
we obtain
\beq
\tilde{\phi}(t) \, = \, \phi\left(x(t)\right) \, = \, t,
\eeq
which can be analytically continued to all the $t$-plane, implying an infinite
radius of convergence.
}:
\beq
\rho_0 \, = \, \log \frac{16}{7} \, \approx \, 0.826679. 
\eeq
In general, for a series expansion centered in $x_0$ with nearest singularities $a$ and $b$
satisfying
\beq
a \, < \, x_0 \, < \, b,
\eeq
the Bernoulli variable $t$ is defined as:
\beq
t \, = \, \log \left( \frac{b - x_0}{x_0 - a} \, \frac{x - a}{b - x} \right).
\eeq
Note that
\beq
t \, = \, 0 \, \iff \, x \, = \, x_0, 
\eeq
while 
\bea
t &\to& - \, \infty ~~~~ {\rm for} ~~~~ x \, \to \, a^+ \, ;
\nn\\ 
t &\to& + \, \infty ~~~~ {\rm for} ~~~~ x \, \to \, b^- .
\eea
The following particular case are relevant in the following:
\begin{itemize}
\item
$x_0 \to \infty$:
\beq
t \, = \, \log \left( \frac{x - a}{x - b} \right) \, ;
\eeq
\item
$b \to \infty$:
\beq
t \, = \, - \, \log \left( \frac{x - a}{x_0 - a} \right) \, ;
\eeq
\item
$a \to - \infty$:
\beq
t \, = \, - \, \log \left( \frac{x - b}{x_0 - b} \right) \, .
\eeq
\end{itemize}

\subsection{Expansion Around $x \, = \, 8$}

In this section we consider the expansion of $F_1(x)$ around the (space-like) point
$x \, = \, 8$, which is a regular singular point of the differential equation not
corresponding to any threshold or pseudothreshold of the diagram.
The singular point of $F_1(x)$ closest to $x = 8$ 
is located in $x \, = \, 0$. We then expect the expansion around $x \, = \, 8$ 
to have a radius of convergence 
\beq
R_8 \, = \, 8, 
\eeq
so that the series converges in the complex $x$-plane for
\beq
|x - 8| \, < \, 8.
\eeq
For real $x$, that means:
\beq
0 \, < \, x \, < \, 16.
\eeq

\subsubsection{Homogeneous Equation}

By solving the indicial equation as in the previous section, we
obtain a double zero in zero, so that the homogeneous equation has 
a solution of the form:
\beq
F_1^{(0)}(x)  =  \sum_{n=0}^{\infty} a_n  (x - 8)^n
 +  \log (x - 8)  \sum_{n=0}^{\infty} b_n  (x - 8)^n.
\eeq
The coefficients are, of course, different from those ones of the
previous section --- we use the same notation to avoid introducing
too many symbols.
The first few coefficients read:
\begin{align}
a_1 &= - \, \frac{5}{24} \, a_0 \, - \, 
\frac{5}{72} \,  b_0 \, ;
&
b_1 &= - \,
\frac{5}{24} \, b_0 \, ;
\\
a_2 &= 
\frac{59}{1728} \, a_0 \, + \, \frac{187}{10368} \, b_0 \, ;
&
b_2 &=
\frac{59}{1728} \, b_0 \, ;
\\
a_3 &=  - \,
\frac{635}{124416} \, a_0 \, - \, \frac{7561}{2239488} \, b_0 \, ;
&
b_3 &= - \,
\frac{635}{124416} b_0  \, ;
\\
a_4 &=
\frac{2171}{2985984} \, a_0 \, + \, \frac{59447}{107495424} \, b_0 \, ;
&
b_4 &= 
\frac{2171}{2985984} \, b_0 \, .
\end{align}
We have again a triangular structure of the coefficients,
as in the previous case.

\subsubsection{Complete Equation}

The expansion of the known term around $x=8$ is of the following form:
\beq
C(x) = \sum_{n=-1}^{\infty} q_n (x - 8)^n \, ,
\eeq
where the first three coefficients $q_n$ are:
\bea
q_{-1} & = & 
            \frac{1}{2304} \pi^2 
          + \frac{1}{768} M_0 \sqrt{2} 
          - \frac{1}{256} M_1 
          + \frac{1}{192} M_2 \sqrt{2}  \log 2
          + \frac{1}{256}  \log^2 2 \, ; 
\\
q_{0} & = & 
          - \frac{13}{82944} \pi^2 
          - \frac{29}{55296} M_0 \sqrt{2} 
          + \frac{13}{9216} M_1 
          - \frac{29}{13824} M_2 \sqrt{2}  \log 2
          + \frac{1}{768}  \log 2 \nn
\\ & & 
          - \frac{13}{9216}  \log^2 2
          - \frac{1}{1024}  \log 3 \, ; 
\\
q_{1} & = & 
            \frac{451}{11943936} \pi^2 
          + \frac{4637}{31850496} M_0 \sqrt{2} 
          - \frac{451}{1327104} M_1 
          + \frac{4637}{7962624} M_2 \sqrt{2}  \log 2 
\nn\\
& & 
          - \frac{551}{884736}  \log 2
          + \frac{451}{1327104}  \log^2 2
          + \frac{61}{147456}  \log 3 \, .
\eea
We have defined the following transcendental constants: 
\bea
M_0   & \equiv & \int_{0}^{1} \frac{\log(1+y)}{\sqrt{y(1+y)}} dy 
\nonumber\\
&=&
\frac{\pi^2}{6} + 4\log 2 \log(\sqrt{2}-1) + 2 \log^2(\sqrt{2}-1)
+ 4 {\rm Li}_2 \left[ i (\sqrt{2}-1)\right]
+ 4 {\rm Li}_2 \left[ - i (\sqrt{2}-1)\right]
\nonumber\\
&\approx& 0.425435  \, , 
\\
M_1 & = & \frac{\pi^2}{6} + \frac{9}{2} \log^2 2 + \mbox{Li}_2 \left( -
\frac{1}{8} \right) \approx 3.68568 \, ;\\
M_2 & = & \log(1+\sqrt{2}) \approx 0.881374 \, .
\eea
The general solution of the inhomogeneous equation reads:
\beq
\bar{F}_1(x)  \, = \, \sum_{n=0}^{\infty} r_n  (x - 8)^n
 +  \log (x - 8)  \sum_{n=0}^{\infty} p_n  (x - 8)^n ,
\eeq
with the first three terms given by:
\bea
r_{1} & = &  
          - \frac{1}{2304} \pi^2 
          - \frac{1}{768} M_0 \sqrt{2} 
          + \frac{1}{256} M_1 
          - \frac{1}{192} M_2 \sqrt{2}  \log 2
          - \frac{1}{256}  \log^2 2
\nn\\
& &   
             - \frac{5}{24} r_0 - \frac{5}{72} p_0      \, ;
\\
p_{1} & = &  - \frac{5}{24} p_0 \, ; 
\\ 
&~&
\nn\\
r_{2} & = & 
            \frac{19}{165888} \pi^2 
          + \frac{79}{221184} M_0 \sqrt{2} 
          - \frac{19}{18432} M_1 
          + \frac{79}{55296} M_2 \sqrt{2}  \log 2
          - \frac{1}{3072}  \log 2 
\nn\\
& &  
          + \frac{19}{18432}  \log^2 2
          + \frac{1}{4096}  \log 3 
          + \frac{59}{1728} r_0 + \frac{187}{10368} p_0 \, ; 
\\
p_{2} & = &  \frac{59}{1728} p_0 \, .
\eea
A particular solution can be obtained by setting $p_0 \, = \, 0$,
which makes all the coefficients $p_n$ vanishing:
that way the series multiplied by the logarithm disappears
from $\bar{F}_1$. One can also set $r_0 \, = \, 0$.  
The first three coefficients are then given by:
\bea
r_0 & = & 0 \, ; 
\\
p_0 & = & 0 \, ; 
\\
&~&
\nn
\\
r_{1} & = &  
          - \frac{1}{2304} \pi^2 
          - \frac{1}{768} M_0 \sqrt{2} 
          + \frac{1}{256} M_1 
          - \frac{1}{192} M_2 \sqrt{2}  \log 2
          - \frac{1}{256}  \log^2 2 \, ;
\\
p_{1} & = &  0 \, ; 
\\ 
&~&
\nn
\\
r_{2} & = & 
            \frac{19}{165888} \pi^2 
          + \frac{79}{221184} M_0 \sqrt{2} 
          - \frac{19}{18432} M_1 
          + \frac{79}{55296} M_2 \sqrt{2}  \log 2
          - \frac{1}{3072}  \log 2 
\nn\\
& &  
          + \frac{19}{18432}  \log^2 2
          + \frac{1}{4096}  \log 3 \, ; 
\\
p_{2} & = &  0 \, .
\eea
The general solution of the differential equation is given by
\beq
F_1(x) \, = \, F_1^{(0)}(x) \, + \, \bar{F}_1(x) \, ,
\eeq
and depends on the arbitrary constants $a_0$ and $b_0$
entering $F_1^{(0)}(x)$. Since we know 
from general arguments that $x \, = \, 8$ is a regular point for the solution, 
we can impose the logarithmic series be absent by requiring
\beq
b_0 \, = \, 0 \, .
\eeq
This condition gives:
\beq
\label{F1+8}
F_1(x) \, = \, \sum_{n=0}^{\infty} s_n (x - 8)^n \, ,
\eeq
where:
\bea
s_0 & = & a_0 \, ; 
\label{freeconst8} \\
s_1 & = & - \frac{5}{24} \, a_0 + r_{1} \, ;\\
s_2 & = & \frac{59}{1728}\,  a_0 + r_{2} \, ,
\eea
etc.. 
Then, the expansion of $F_1(x)$ around $x \, = \, 8$ given in eq.~(\ref{F1+8}) 
does not determine $F_1(x)$ uniquely, because it still contains the free parameter $a_0$.
By using the matching procedure described in sect.~{\ref{match1}}, we obtain the
following numerical estimate for this coefficient:
\beq
a_0 \, = \, 0.0321062814000779405116 \, . 
\eeq

\subsubsection{Improved Expansion}

In order to improve the convergence of the series so far obtained, we move
from the series in $x$ to the one in the Bernoulli variable
\beq
t \, \equiv \, \log \frac{x}{8} \, ,
\eeq
with the inverse:
\beq
x \, = \, 8 e^t \, = \, 8 + 8t + 4t^2 + \frac{4}{3} t^3 + \frac{1}{3} t^4 
+ \frac{1}{15} t^5 + \frac{1}{90} t^6 + {\mathcal O}(t^7) \, .
\label{bernoulli8}
\eeq
Unlike the previous case ($x = 0$), the above series has an infinite radius
of convergence:
\beq
r_8 \, = \, \infty \, .
\eeq
Substituting the r.h.s. of the above equation
in the series expansion for $F_1(x)$ obtained previously 
and, finally, expanding in powers of $t$, we have:
\beq
\tilde{F}_1(t) \, \equiv \, F_1\left(x(t)\right) \, = \,
\sum_{n=0}^{\infty} \alpha_n \, t^n \, ,
\eeq
where the first three coefficients $\alpha_n$ are:
\bea
\alpha_1 &=&  8 \, s_0 \, ;
\\
\alpha_2 &=& 
            \frac{73}{54} a_0
          + \frac{29}{5184} \pi^2 
          + \frac{61}{3456} M_0 \sqrt{2} 
          - \frac{29}{576} M_1 
          + \frac{61}{864} M_2 \sqrt{2}  \log 2
          - \frac{1}{48}  \log 2 
\nn\\
& & 
          + \frac{29}{576}  \log^2 2
          + \frac{1}{64}  \log 3 \, ;
\\
\alpha_3 &=& 
          - \frac{343}{486} a_0
          - \frac{203}{46656} \pi^2 
          - \frac{929}{62208} M_0 \sqrt{2} 
          + \frac{203}{5184} M_1 
          - \frac{929}{15552} M_2 \sqrt{2}  \log 2 
\nn\\
& &
          + \frac{7}{192}  \log 2
          - \frac{203}{5184}  \log^2 2
          - \frac{7}{288}  \log 3 \, .
\eea
A computation of the convergence radius $\rho_8$ of the series in $t$
analogous to the one of the previous section gives:
\beq
\rho_8 \, = \, \log 2 \, \approx \,  0.693147.
\eeq

\subsubsection{The Matching Condition}
\label{match1}
$a_0$ can be computed by imposing that the series in 
eq.~(\ref{Eq0}), which is completely determined, and that one in eq.~(\ref{F1+8}) assume 
the {\it same} value in a given point in the intersection of the respective domains of 
convergence.
One has to take a point lying in the interval 
\beq
- \, 1 \, < \, x \, < \, 1,
\eeq 
where the series of eq.~(\ref{Eq0}) converges, 
as well as in the interval
\beq
0 \, < \, x \, < \, 16,
\eeq
where the series in eq.~(\ref{F1+8}) converges.
One has therefore to choose a point in the interval
\beq
\label{intersect}
0 \, < \, x \, < \, 1,
\eeq
such as for example $ x \, = \, 1/2$. 
If we deal with infinite series, this procedure exactly
determines the coefficient $a_0$.
As we have shown above, however, we can only determine an arbitrary, but finite,
number of coefficients of both serieses and the matching has to be made 
in an approximate numerical way by using truncated series.
The number of terms of the two serieses that must be computed depend on the 
required precision on $a_0$.
Our goal is to give $F_1(x)$ with a relative precision 
of better then $10^{-15}$ (double precision) using a relatively small number 
of terms in the serieses (around 30). 
The problem is that any point in the interval (\ref{intersect}) 
is close to the boundary of the convergence domain for the series centered around 
$x \, = \, 8$, where convergence is slow, implying that many terms are needed for high accuracy.
In other words, a direct matching between the series in 
eq.~(\ref{Eq0}) and that one in eq.~(\ref{F1+8}) is not the good algorithm. 
As we have shown before, a first improvement is obtained by re-writing the
series expansions in terms of the relevant Bernoulli variables.
But it is convenient to use additional series expansions centered around auxiliary 
regular points in the range $0 \, < \, x \, < \, 8$, such as for example $x \, = \, 3$.
This topic will be discussed in more detail in sect.~\ref{AddMat}.

\subsection{Expansion Around $x \, = \, \infty$ --- Large Momentum Expansion}

In this section we consider the expansion of $F_1(x)$ around $x \, = \, \infty$,
which generates a large-momentum expansion.
Since the closest singularity to $x \, = \, \infty$ is at $x \, = \, 8$,
we expect the expansion around infinity to be convergent outside the
circle of radius 8, i.e. for
\beq
\label{infinity}
| x |  \, > \, 8.
\eeq
The expansion at infinity is studied systematically by changing variable to
\beq
y  \equiv  \frac{1}{x}
\eeq
and considering the limit 
\beq
\label{yto0}
y  \to  0.
\eeq

\subsubsection{Homogeneous Equation}

Let us write the solution as usual as:
\beq
F_1^{(0)}(y)  =  S(y)  \sum_{n=0}^{\infty} A_n  y^n.
\eeq
The pre-factor is assumed to have power-like form,
\beq
S(y)  =  y^{\beta} ,
\eeq
and to be a solution of the homogeneous equation in the limit (\ref{yto0}):
\beq
S''(y)  -  \frac{3}{y}  S'(y)  +  \frac{4}{y^2}  S(y)  =  0 \, .
\eeq
The indicial equation is $(\beta-2)^2 = 0$, with a double zero in $\beta = 2$ and 
therefore the solution is, in the original $x$ variable, of the form:
\beq
F_1^{(0)}(x)  =  \sum_{n=2}^{\infty} \frac{a_n}{x^n}
 +  \log x  \sum_{n=2}^{\infty} \frac{b_n}{x^n} \, .
\eeq
Because of the presence of the logarithmic term in front
of the second series, the infinity is a singular point
--- more exactly, a branch point of infinite order.
Furthermore, since positive powers of $x$ do not appear, $x\, = \, \infty$ 
is a regular point for the serieses above. 
Let us write a bunch of coefficients in terms of the
lowest-order ones $a_2$ and $b_2$:
\begin{align}
a_3 &= 2 \, a_2  -  3 \, b_2 \, ;
&
b_3 &= 2 \, b_2 \, ;
\\
a_4 &=  10 \, a_2  -  \frac{33}{2} \, b_2 \, ;
&
b_4 &= 10 \, b_2 \, ;
\\
a_5 &=  56 \, a_2  -  100 \, b_2 \, ;
&
b_5 &= 56 \, b_2 \, ;
\\
a_6 &= 346 \, a_2  -  \frac{2561}{4} \,  b_2 \, ;
&
b_6 &= 346 \, b_2 \, ;
\\
a_7 &= 2252 \, a_2  -  \frac{42631}{10} \,  b_2 \, ;
&
b_7 &= 2252 \, b_2 \, .
\end{align}
Let us notice the power-like growth of the coefficients,
which should scale asymptotically as $\approx 8^n$
($b_7/b_6$ is already $\approx 6.5$).

\subsubsection{Complete Equation}

The non-trivial part of the expansion of the known
term $C(x)$ around $x=\infty$ is related to the expansion
of the GHPL's around this point, which is discussed in appendix \ref{appb}.
The expansion is of the form:
\beq
C(x) =  \sum_{n=4}^{\infty} \frac{k_n}{x^n}
 +  \log x  \sum_{n=4}^{\infty} \frac{l_n}{x^n}
 +  \log^2 x  \sum_{n=4}^{\infty} \frac{m_n}{x^n}
~~~~~~~~~~~~ (|x|>8),
\eeq
where the lowest-order coefficients read:
\begin{align}
k_4 &= \frac{\pi ^2}{48}  \, ;
&
l_4 &= 0 \, ;
&
m_4 &= - 
\frac{7}{16} \, ;
\\
k_5 &= 
\frac{1}{2}  +  \frac{13 \pi ^2}{48} \, ;
&
l_5 &= - 
\frac{7}{4} \, ;
&
m_5&  = - 
\frac{43}{16} \, ;
\\
k_6 &= 
\frac{1}{8}  +  \frac{125 \pi ^2}{48} \, ;
&
l_6 &= - 
\frac{131}{8} \, ;
&
m_6 &= - 
\frac{331}{16} \, ;
\\
k_7 &=  - 
\frac{40}{9}  +  \frac{357 \pi ^2}{16} \, ;
&
l_7 &= - 
\frac{3437}{24} \, ;
&
m_7 &= - 
\frac{2569}{16} \, ;
\\
k_8 &=  - 
\frac{18323}{288} + \frac{8827 \pi ^2}{48} \, ;
&
l_8 &= - 
\frac{56953}{48} \, ;
&
m_8 &= - 
\frac{20301}{16} \, .
\end{align}
Since the differential equation involves a second derivative
and the known term has double-logarithmic terms, the solution
must contain up to four powers of the logarithm:
\beq
\bar{F}_1(x) =  \sum_{n=2}^{\infty} \frac{p_n}{x^n}
 +  \log x  \sum_{n=2}^{\infty} \frac{q_n}{x^n}
 +  \log^2 x  \sum_{n=2}^{\infty} \frac{r_n}{x^n}
 +  \log^3 x  \sum_{n=2}^{\infty} \frac{s_n}{x^n}
 +  \log^4 x  \sum_{n=2}^{\infty} \frac{t_n}{x^n} \, .
\eeq
Substituting the above form for the solution into the
equation, we obtain the following relations among the
coefficients:
\bea
r_2 &=& -  \frac{\pi^2}{96} \, ;
\\
s_2 &=& 0 \, ;
\\
t_2 &=& \frac{7}{192} \, ;
\\
&~&
\nn
\\
p_3 &=& 2 \, p_2  -  3 \, q_2  -  \frac{5}{48}  \pi ^2
 +  \frac{27}{8} \, ;
\\
q_3 &=& 2 \, q_2 + \frac{\pi ^2}{16}  +  \frac{9}{8} \, ;
\\
r_3 &=&  -  \frac{13}{16}  -  \frac{\pi^2}{48} \, ;
\\
s_3 &=&  -  \frac{7}{16} \, ;
\\
t_3 &=&  \frac{7}{96} \, .
\eea
The values of $p_2$ and $q_2$ are left undetermined by the differential equation
and a particular solution can be found by imposing for instance 
\beq
\label{evidenziata}
p_2 \, = \, 0 ; ~~~~~~ q_2 \, = \, 0 ,
\eeq
which we assume from now on. The general solution is given by:
\bea
\label{Eq+oo}
F_1(x) & = & \, F_1^{(0)}(x) + \bar{F}_1(x) 
\\
& = & \, \sum_{n=2}^{\infty} \frac{\tilde{p}_n}{x^n}
 +  \log x  \sum_{n=2}^{\infty} \frac{\tilde{q}_n}{x^n}
 +  \log^2 x  \sum_{n=2}^{\infty} \frac{r_n}{x^n}
 +  \log^3 x  \sum_{n=2}^{\infty} \frac{s_n}{x^n}
 +  \log^4 x  \sum_{n=2}^{\infty} \frac{t_n}{x^n},
\nonumber
\eea
where we have defined: 
\beq
\tilde{p}_n \, \equiv \, p_n + a_n \, ; 
~~~~~~~~
\tilde{q}_n \, \equiv \, q_n + b_n.
\eeq
Let us note that the coefficients of the double, triple
and fourth logarithm are uniquely determined because
the associated homogeneous equation has no terms of this form:
we therefore obtain exact analytic expressions for these coefficients.
On the contrary, the coefficients of the terms with the single logarithm
and of the terms without logarithm are determined up to 
a solution of the homogeneous equation, and then they do
depend on the two arbitrary constants 
\bea
\tilde{p}_2 ~~~~ \mbox{and} ~~~~ \tilde{q}_2 \, .
\label{freeconstoo}
\eea
The latter can be found by matching the 
values of the series in eq.~(\ref{Eq+oo}) with that one in eq.~(\ref{F1+8}) in a 
point in the range 
\beq
8 \, < \, x \, < \, 16. 
\eeq
By following the matching procedure outlined in the previous section and described 
in detail in sec.~\ref{AddMat}, we obtain
\bea
\tilde{p}_2 & = & - \, 1.04850063265766512303 \, ; 
\label{infty1} \\
\tilde{q}_2 & = & + \, 1.50257112894949285675 \, .
\label{infty2} 
\eea
By decomposing the two-loop integral into different infrared and ultraviolet regions,
a large momentum expansion of $F_1$ has been derived in 
ref.~\cite{Smirnov:1998vk}, which reads in our normalization containing an additional
factor $1/16$ : 
\beq
F_1(x) \, = \, \frac{1}{x^2}
\left(
\frac{7}{192} \log^4 x
\, - \, \frac{\pi^2}{96} \, \log^2 x
\, + \, \frac{5}{4} \, \zeta(3) \, \log x
\, - \, \frac{31}{2880} \, \pi^4
\right)
\, + \, O\left(\frac{1}{x^3}\right),
\eeq
where $\zeta(3) \equiv \sum_{n=1}^{\infty} 1/n^3 = 1.20206 \cdots$.
By comparing with the first-order terms of our complete solution, 
we find an analytical agreement in the coefficient of the fourth,
triple and double logarithm\footnote{ We thank V. Smirnov for confirming a typo 
in the coefficient of the $\log^2 x$ term in \cite{Smirnov:1998vk}, which must 
be divided by a factor 3.}.
As far as the numerically determined coefficients are concerned,
differences with the analytical expressions are at most $O\left(10^{-19}\right)$,
well within the expected numerical uncertainty.
We can therefore use in place of our numerical estimates above the values:
\beq
\tilde{p}_2 \, = \, - \, \frac{31}{2880} \, \pi^4,
~~~~~
\tilde{q}_2 \, = \, + \, \frac{5}{4} \, \zeta(3).
\eeq
Our final expression for the large-momentum expansion of the amplitude $F_1(x)$
of the crossed ladder diagram is that one in eq.~(\ref{Eq+oo}), 
with the first coefficients given by:
\bea
\tilde{p}_2 &=&  - \, \frac{31}{2880} \, \pi^4 \, ;
\\
\tilde{q}_2 &=& + \, \frac{5}{4} \, \zeta(3) \, ;
\\
r_2 &=& -  \, \frac{\pi^2}{96} \, ;
\\
s_2 &=& 0 \, ;
\\
t_2 &=& \frac{7}{192} \, ;
\\
&~&
\nn
\\
\label{ps1}
\tilde{p}_3 &=& 
   \frac{27}{8} 
 - \frac{5}{48} \, \pi^2
 - \frac{31}{1440} \, \pi^4
 - \frac{15 }{4} \, \zeta(3) \, ;
\\
\tilde{q}_3 &=& 
   \frac{9}{8}
 + \frac{\pi ^2}{16}
 + \frac{5}{2} \, \zeta(3) \, ;
\\
r_3 &=&  -  \frac{13}{16}  -  \frac{\pi^2}{48} \, ;
\\
s_3 &=&  -  \frac{7}{16} \, ;
\\
\label{ps2}
t_3 &=&  \frac{7}{96} \, .
\eea
Let us remark that eqs.~(\ref{ps1}-\ref{ps2}) provide analytic
expressions of the coefficients of the leading power-suppressed terms in the
large momentum expansion.

\subsubsection{Improved Expansion}

In order to accelerate the convergence of the series so far obtained, we move
from the series in $x$ to the one in the Bernoulli variable
\beq
t = \log \frac{x-8}{x+1} \, ,
\eeq
with the inverse:
\beq
\frac{1}{x} = \frac{1-e^t}{8+e^t} = - \frac{1}{9} t - \frac{7}{162} t^2
- \frac{11}{1458} t^3 + \frac{35}{52488} t^4 + 
  \frac{1733}{2361960} t^5 + \frac{7217}{42515280} t^6 
+ {\mathcal O}(t^7) \, .
\label{bernoullioo}
\eeq
The convergence radius of the above series is equal to the one of the
Bernoulli variable for the expansion around $x \, = \, 0$:
\beq
r_{\infty} \, = \, \sqrt{\pi^2 + 9 \log^2 2}.
\eeq
Substituting the r.h.s. of the above equation
in the series expansion for $F_1(x)$ obtained previously 
and finally expanding in powers of $t$, we have:
\beq
\tilde{F}_1(t) \, = \, \sum_{n=2}^{\infty} \alpha_n t^n
 +  \log x  \sum_{n=2}^{\infty} \beta_n t^n
 +  \log^2 x  \sum_{n=2}^{\infty} \gamma_n t^n
 +  \log^3 x  \sum_{n=2}^{\infty} \delta_n t^n
 +  \log^4 x  \sum_{n=2}^{\infty} \phi_n t^n \, ,
\eeq
where the first three coefficients $\alpha_n$, $\beta_n$, $\gamma_n$, 
$\delta_n$, and $\phi_n$ are: 
\begin{eqnarray}
\alpha_2 &=& \, \frac{1}{81} \, a_2 \, ; 
\\
\beta_2  &=& \, \frac{1}{81} \, b_2 \, ;
\\
\gamma_2 &=& \, \frac{1}{7776} \pi^2 \, ; 
\\
\delta_2 &=& \, 0 \, ;
\\
\phi_2   &=& \, \frac{7}{15552} \, ; 
\\
&\,& 
\nn\\
\alpha_3 &=& \, 
            \frac{5}{34992} \pi^2
          - \frac{1}{216}
          + \frac{1}{243} b_2
          + \frac{5}{729} a_2 \, ;
\\
\beta_3  &=& \, 
          - \frac{1}{11664} \pi^2
          - \frac{1}{648}
          + \frac{5}{729} b_2 \, ; 
\\
\gamma_3 &=& \, 
          - \frac{5}{69984} \pi^2
          + \frac{13}{11664} \, ;
\\
\delta_3 &=& \, \frac{7}{11664} \, ; 
\\
\phi_3   &=& \, \frac{35}{139968} \, ;
\\
&\,& 
\nn\\
\alpha_4 &=& \, 
            \frac{95}{1259712} \pi^2
          - \frac{2659}{839808}
          + \frac{5}{2187} b_2
          + \frac{49}{26244} a_2 \, ; 
\\
\beta_4  &=& \, 
          - \frac{5}{104976} \pi^2
          - \frac{29}{46656}
          + \frac{49}{26244} b_2 \, ;
\\
\gamma_4 &=& \, 
          - \frac{49}{2519424} \pi^2
          + \frac{125}{139968} \, ; 
\\
\delta_4 &=& \, \frac{35}{104976} \, ;
\\
\phi_4   &=& \, \frac{343}{5038848} \, ,
\end{eqnarray}
where (see eq.~(\ref{evidenziata}))
\beq
a_2 \, = \, \tilde{p}_2 \, = \, - \, \frac{31}{2880} \, \pi^4 ;
~~~~~~~~~
b_2 \, = \, \tilde{q}_2 \, = \, + \, \frac{5}{4} \, z(3) .
\eeq
The convergence radius of the expansion in $t$ is:
\beq
\rho_{\infty} \, = \, \log \frac{16}{7} \, \approx \, 0.826679 \, .
\eeq

\subsubsection{Analytic Continuation to $x \, = \, - \, \infty$ }

The expansion of the amplitude $F_1(x)$ for large time-like momenta,
namely for (cfr. eq.~(\ref{infinity}))
\beq
- \, \infty \, < \, x \, < \, - \, 8, 
\eeq
can be found from the asymptotic expansion in the space-like region ($x \, > \, 8$) 
simply by analytic continuation.
With the causal prescription
\beq
x \, \to \, - | x | - i \epsilon \, ,
\eeq
we have in addition to the replacements (\ref{prime2}):
\bea
\log^3 x & \to & \log^3 |x|  - 3 \pi ^2 \log |x| - 3 i \pi \log^2 |x| + i \pi ^3 \, ;
\nn\\
\log^4 x & \to &
\log^4 |x| - 6 \pi ^2 \log^2 |x|  + \pi ^4 - 4 i \pi \log^3 |x|  + 4 i \pi ^3 \log |x| \, .
\eea
$F_1(x)$ then becomes complex for $x \, < \, 0$ with: 
\bea
{\rm Re} \, F_1(x) & = & \,  \sum_{n=2}^{\infty} \frac{\tilde{p}_n}{x^n}
 +  \log |x|  \sum_{n=2}^{\infty} \frac{\tilde{q}_n}{x^n}
 +  (\log^2 |x| - \pi^2)  \sum_{n=2}^{\infty} \frac{r_n}{x^n} \, +
\\
&+& \left( \log^3 |x| - 3 \pi^2 \log |x| \right)  \sum_{n=2}^{\infty} \frac{s_n}{x^n}
 +  \left( \log^4 |x| - 6 \pi^2 \log^2 |x| + \pi^4 \right)  
\sum_{n=2}^{\infty} \frac{t_n}{x^n} \, ;
\nn \\
{\rm Im} \, F_1(x) & = & \, \pi 
\Bigg[ 
 - \sum_{n=2}^{\infty} \frac{\tilde{q}_n}{x^n}
 - 2 \log |x|  \sum_{n=2}^{\infty} \frac{r_n}{x^n}
 - \left( 3 \log^2 |x| - \pi^2 \right)  \sum_{n=2}^{\infty} \frac{s_n}{x^n} \, +
\nn\\
&-& \left( 4 \log^3 |x| - 4 \pi^2 \log |x| \right) \sum_{n=2}^{\infty} \frac{t_n}{x^n}
\Bigg] \, .
\eea 

\subsection{Expansion Around $x \, = \, - \, 1$}

In this section we consider the expansion around $x \, = \, - \, 1$,
the only singularity of the differential equation in the time-like region.
Since the nearest singularity to $x = - 1$ is the origin $x = 0$,
we expect the convergence radius of the expansion around this 
point to be 
\beq
R_{-1} \, = \, 1, 
\eeq
i.e. convergence in the circle
\beq
|x + 1| \, < \, 1.
\eeq
On the real axis, that means:
\beq
- \, 2 \, < \, x \, < \, 0.
\eeq

\subsubsection{Homogeneous Equation}

By solving the indicial equation, the solution
turns out to be of the form:
\beq
F_1^{(0)}(x) \, = \, \sum_{n=0}^{\infty} a_n \, (x \, + \, 1)^n
\, + \, \log (x \, + \, 1) \, \sum_{n=0}^{\infty} b_n \, (x \, + \, 1)^n.
\eeq
The differential equation allows one to express all the coefficients
of the expansion above in terms of two of them, such as for
example the lowest-order ones $a_0$ and $b_0$:
\begin{align}
a_1 &= \frac{4}{3} a_0 + \frac{4}{9} b_0 \, ;
&
b_1& =
\frac{4}{3}  b_0 \, ;
\\
a_2 &= 
\frac{41 }{27} a_0 + \frac{62}{81}  b_0 \, ;
&
b_2 &=
\frac{41 }{27} b_0 \, ;
\\
a_3 &= 
\frac{400 }{243} a_0 + \frac{2200 }{2187} b_0 \, ;
&
b_3 &=
\frac{400 }{243} b_0 \, .
\end{align}
We have a triangular structure of the coefficients, as in the
case of the expansion around $x \, = \, 0$.

\subsubsection{Complete Equation}

The expansion of the known term around $x \, = \, - \, 1$ is of the form:
\beq
C(x)  \, = \, \sum_{n=-1}^{\infty} q_n  (x  +  1)^n 
 +  \log(x+1) \sum_{n=0}^{\infty} r_n  (x  +  1)^n \, . 
\eeq
The leading singularity for $x \to - \, 1$ is therefore a single pole
and the subleading singularity is a logarithmic one related to
the second series.
The first few coefficients read:
\bea
q_{-1} &=& 
\frac{\pi^2}{24} 
 +  \frac{\sqrt{5}}{24}  K
 -  i  \frac{\sqrt{5}}{12}  \pi  K' \, ; 
\\
&\,&
\nn\\
q_0 &=& 
-  \frac{1}{4}  +  \frac{19}{216}  \pi^2
 +  \frac{101}{1080}  \sqrt{5}  K
 +  i \pi \left( \frac{7}{72}  -  \frac{101}{540}  \sqrt{5} 
 K' \right)  \, ;
\\
r_0 &=& \frac{1}{4} \, ; 
\\
&\,&
\nn\\
q_1 &=& 
-  \frac{2}{3}  +  \frac{131}{972}  \pi^2
 +  \frac{1861}{12150}  \sqrt{5}  K
 +  i \pi \left( \frac{1627}{6480}  -  \frac{1861}{6075}  \sqrt{5} 
 K' \right) \, ; 
\\
r_1 &=& \frac{47}{72}  \, ; 
\\
&\,&
\nn\\
q_2 &=& 
- \frac{5177}{4320} + \frac{1589}{8748}  \pi^2
 +  \frac{60178}{273375}  \sqrt{5}  K
 +  i \pi \left( \frac{129527}{291600}  -  \frac{120356}{273375}  \sqrt{5} 
 K' \right)  \, ; 
\\
r_2 &=& \frac{749}{648}  \, ,
\eea
where $K$ and $K'$ are two real transcendental constants
defined as:
\bea
K & \equiv & \int_0^1 \frac{\log y}{\sqrt{y(y \, + \, 4)}} dy \, = \, - \, \frac{\pi^2}{5}
\, \approx \, - \, 1.97392 ;
\\
K' &=& \log \left( \frac{1 \, + \, \sqrt{5}}{2} \right) \, \approx \, 0.481212.
\eea
Let us note that the coefficients above have an imaginary part
related to the 2 massless particles threshold in $s \, = \, 0$.
Unlike previous cases, the expansion point lies indeed in the time-like region.
\noindent
The particular solution of the inhomogeneous equation is of the same form as the 
homogeneous one:
\beq
\bar{F}_1(x) \, = \, \sum_{n=0}^{\infty} s_n  (x  +  1)^n 
 +  \log(x+1) \sum_{n=0}^{\infty} t_n  (x  +  1)^n \, .
\eeq
For the coefficients, we find:
\bea
s_1 & = & \frac{4}{3} s_0 + \frac{4}{9} t_0 + q_{-1} \, ; 
\\
t_1 & = & \frac{4}{3} t_0 \, ;
\\
&~&
\nn\\
s_2 & = & \frac{41}{27} s_0 + \frac{62}{81} t_0 - \frac{1}{8}
+ \frac{59}{864} \pi^2 + \frac{301}{4320} \sqrt{5}  K
+  i \pi \left( \frac{7}{288} - \frac{301}{2160} \sqrt{5} \right) \, ; 
\\
t_2&  = & \frac{1}{16} + \frac{41}{27} t_0 \, ;
\\
&~&
\nn\\
s_3 & = & 
 \frac{400}{243} s_0 + \frac{2200}{2187} t_0 - \frac{13}{54}
+ \frac{169}{1944} \pi^2 + \frac{4381}{48600} \sqrt{5}  K
+  i \pi \left( \frac{313}{6480} - \frac{4381}{24300} \sqrt{5} \right) \, ;
\\
t_3 & = & \frac{1}{8} + \frac{400}{243} t_0 \, ;
\eea
etc., and we can choose $s_0 \, = \, t_0 \, = \, 0$.
We assume therefore from now on the following expressions of the 
coefficients:
\bea
s_0 &=& 0 \, ;
\\
t_0 &=& 0 \, ;
\\
&~&
\nn\\
s_1 & = & \frac{\pi^2}{24} 
 +  \frac{\sqrt{5}}{24}  K
 -  i  \frac{\sqrt{5}}{12}  \pi  K'  \, ; 
\\
t_1 & = & 0 \, ; 
\\
&~&
\nn
\\
s_2 & = & - \frac{1}{8} + \frac{59}{864} \pi^2 + \frac{301}{4320} \sqrt{5}  K
+  i \pi \left( \frac{7}{288} - \frac{301}{2160} \sqrt{5} \right) \, ;
\\
t_2&  = & \frac{1}{16} \, ; 
\\
&~&
\nn
\\
s_3 & = & - \frac{13}{54}
+ \frac{169}{1944} \pi^2 + \frac{4381}{48600} \sqrt{5}  K
+  i \pi \left( \frac{313}{6480} - \frac{4381}{24300} \sqrt{5} \right) \, ;
\\
t_3 & = & \frac{1}{8} \, .
\eea
Finally, the general solution is given by:
\bea
F_1(x) & = & F_1^{(0)}(x) + \bar{F}_1(x) \\
& = & \sum_{n=0}^{\infty} \tilde{s}_n  (x  +  1)^n 
 +  \log(x+1) \sum_{n=0}^{\infty} \tilde{t}_n  (x  +  1)^n \, ,
\label{Eq-1}
\eea
where 
\beq
\tilde{s}_n \, = \, a_n \, + \, s_n ~~~~~ {\rm and} ~~~~~ \tilde{t}_n \, = \, b_n + t_n. 
\eeq
Eq. (\ref{Eq-1}) depends on the two arbitrary constants, 
\beq
a_0 ~~~~ \mbox{and} ~~~~ b_0 \, ;
\label{freeconst-1}
\eeq
that can be determined in a numerical way as shown above:
we equate the series expansions centered
around $x \, = \, 0$ and $x \, = \, - \, 1$, together with their derivatives,
in one point where both series converge, such as for example
$x \, = \, - \, 1/2$.
That way we obtain:
\bea
a_0 & = & - \, 0.07572639563476980715 \, + \, i \, 0.3122156449500221544 \, ; 
\\
| b_0 | & < & 10^{-18} \, ,
\eea
naturally implying 
\beq
\label{naturale}
b_0 \, = \, 0.
\eeq
Let us observe that for $x \, < \, - \, 1$ the logarithm above acquires an imaginary
part related to the 3-particle threshold at $s\, = \, m^2$.
Then, absorptive contributions related to both thresholds formally come out 
as imaginary parts of the coefficients of the serieses and, 
via analytic continuation, of the logarithmic prefactor.

The vanishing of the coefficient $b_0$ of the $\log(1+x)$ term in the expansion
of $F_1$ (eq.~(\ref{naturale})) can be understood as follows.
With a proper routing, $F_1$ can be written as:
\beq
F_1 \, = \, \int \frac{ \mathcal{D}^4 k_1 \mathcal{D}^4 k_2 }{ (q+k_1)^2 + m^2 }
\frac{1}{ (q+k_2)^2 + m^2 }
\frac{1}{ (p_1 + k_1)^2 }
\frac{1}{ (p_2 + k_2)^2 }
\frac{1}{ (p_1 + k_1 - k_2)^2 }
\frac{1}{ (p_2 - k_1 + k_2)^2 } \, .
\eeq
By going to the threshold point $q^2 = - \, m^2$ and neglecting quadratic terms 
in the loop momenta $\sim k_1^2, \, k_2^2, \, k_1\cdot k_2$, 
one obtains in the soft limit:
\beq
F_1 \approx \int \mathcal{D}^4 k_1 \mathcal{D}^4 k_2
~\frac{ 1 }{ q \cdot k_1 }
~\frac{1}{ q \cdot k_2  }
~\frac{1}{ p_1 \cdot k_1 }
~\frac{1}{ p_2 \cdot k_2 }
~\frac{1}{ p_1 \cdot k_1 - p_1 \cdot k_2 }
~\frac{1}{ - p_2 \cdot k_1 + p_2 \cdot k_2 }.
\eeq
The integral above is convergent by power counting in the soft region.
It is also easy to see that there cannot be any collinear singularity.
To have collinear singularities, one should have at least one ``all massless''
3-point vertex, i.e. a 3-point vertex connecting massless 
propagators and/or external lines with light-cone momenta only,
which is not the case.
Since $F_1$ has not any infrared singularity in the point $q^2 = - m^2$,
it is finite for $x \to -1$\footnote{
$F_1$ is ultraviolet finite, as shown previously. 
}
and therefore cannot contain any $\log(1+x)$ 
term in the expansion, implying $b_0 = 0$.

\subsubsection{Improved Expansion}

In order to improve the convergence of the series so far obtained, we move
from the series in $x$ to the one in the Bernoulli variable
\beq
t \, \equiv \, - \, \log (-x) \, ,
\eeq
with inverse:
\beq
x \, = \, - \, e^{-t} \, = \, - 1 + t - \frac{1}{2} t^2 + \frac{1}{6} t^3 - \frac{1}{24} t^4 
+ \frac{1}{120} t^5 - \frac{1}{720} t^6 
+ {\mathcal O}(t^7) \, .
\label{bernoulli-1}
\eeq
The convergence radius of the series above is, as well known, infinite:
\beq
r_{-1} \, = \, \infty \, .
\eeq
Substituting the r.h.s. of the above equation
in the series expansion for $F_1(x)$ obtained previously 
and finally expanding in powers of $t$, we have:
\beq
\tilde{F}_1(t) \, \equiv \, F_1(x(t)) \, = \, \sum_{n=0}^{\infty} \alpha_n t^n
 \, + \, \log (x+1)  \sum_{n=0}^{\infty} \beta_n t^n \, ,
\eeq
where the first three coefficients $\alpha_n$ and $\beta_n$ are: 
\begin{align}
\alpha_0 & = a_0 \, ; & \beta_0  & = 0 \, ; \\
\alpha_1 & = \frac{4}{3} a_0 
 +  \frac{\pi^2}{24} 
 +  \frac{\sqrt{5}}{24}  K
 -  i  \frac{\sqrt{5}}{12}  \pi  K' \, ; &
\beta_1  & = 0 \, ; \\
\alpha_2 & = 
- \frac{1}{8}
+ \frac{41}{864} \pi^2 + \frac{211}{4320} \sqrt{5}  K
+  i \pi \left( \frac{7}{288} - \frac{211}{2160} \sqrt{5} \right)  \, ; &
\beta_2 & =  \frac{1}{16}  \, .
\end{align}
The convergence radius of the expansion in $t$ is given by:
\beq
\rho_{-1} \, = \, \log 2 \, \approx \, 0.693147 \, .
\eeq

\subsection{Series Expansions Around Regular Points and Additional Matching Points}
\label{AddMat}

In the previous sections we illustrated the method of the solution of the differential
equation on the MI $F_1(x)$, building series around the singular points only. 
That is certainly necessary but not sufficient to determine $F_1$ on the entire
real axis.
The expansions around all the singular points do not 
allow indeed to compute the master integral in the range
\beq
\label{uncovered}
-\, 8 \, < \, x \, < \, - \, 2.
\eeq
To cover this region, it is necessary to perform expansions
of $F_1(x)$ around auxiliary (regular) points $x_i$, expansions which are
not multiplied by any singular function:
\beq
F_1(x) \, = \, \sum_{n=0}^{\infty} c_n^{(i)} \, (x - x_i)^n.
\eeq
It is clear that each new series expansion carries two unknown coefficients,
which have to be determined by matching with known serieses, as we have
made for the singular points.
To cover the region (\ref{uncovered}), we have constructed two additional series 
expansions centered in 
\beq
x \, = \, - \, 8 ~~~~~ {\rm and} ~~~~~ x \, = \, - \, 3. 
\eeq
Even though the space-like region $0 \, < \, x \, < \, 8$ is in principle covered
by the expansions around the singular points,
we have found it convenient to consider also series expansions
centered in 
\beq
x \, = \, 1 ~~~~~ {\rm and} ~~~~~ x \, = \, 3.
\eeq 
The matching procedure is the following\footnote{
The practical advantage of this method with respect to the general analytic continuation 
by (partially overlapping) circles is related to the use of the differential equation
to generate both serieses. 
}:
\begin{enumerate}
\item
we match the series centered in $x \, = \, 1$ with the series centered in $x \, = \, 0$, 
which is completely determined. That way we fix the two arbitrary coefficients
entering the expansion in $x \, = \, 1$;
\item
we then match the series in $x \, = \, 3$ with the one in $x \, = \, 1$;
\item
we finally match the series in $x \, = \, 8$ with
the series in $x \, = \, 3$. 
\end{enumerate}
With the above method, we have been able to cover the range $0 \, < \, x \, < \, 13$ 
with a numerical precision of about 21 digits. 
In the range $8 \, < \, x \, < \, \infty$, we need an
additional series expansion in $x \, = \, 16$, to be
matched with the series centered in $x \, = \, 8$.

It is remarkable that the 9 expansion points above (4 singular + 5 regular ones) 
are sufficient to build a routine evaluating numerically the 
function $F_1(x)$ for every value of the variable $x$ in the real axis 
with a relative precision of about $10^{-20}$.

The internal consistency of the method can be checked as follows.
The series expansion centered in $x \, = \, 0$ is completely determined by the
initial conditions that we have been able to find. Matching the free
parameters of the series expansions centered in $x \, = \, - \, 1$ and $x \, = \, 1$ 
with the one centered in $x \, = \, 0$ we obtain two other series without free parameters. 
Now we move along the real positive axis up to $x \, = \, \infty$; 
\beq
x = 0 ~~~ \rightarrow ~~~ x = 1 ~~~ \rightarrow ~~~ x = 3 ~~~
\rightarrow ~~~ x = 8 ~~~ \rightarrow ~~~ x = \infty \, . 
\eeq
After that, we perform an analytic
continuation in order to find the series expansion in $x \, = \, - \, \infty$ and, 
finally, we reach the series expansion in $x \, = \, - \, 1$. What we find is a perfect
agreement, within the required precision, between the free parameters that 
we fixed matching the series expansion in $x \, = \, - \, 1$ with the one in $x \, = \, 0$ and 
those found performing the several matches along the real axis, back to
$x \, = \, - \, 1$
\footnote{
Since we deal in general with multivalued functions, one has to be careful to remain within
the same sheet of the Riemann surface by using consistently the causal $+i\epsilon$ prescriptions.
}.
Another (independent) check of our matching procedure is provided 
by the values of the coefficients in eqs.~(\ref{infty1},\ref{infty2}),
which are in agreement with the asymptotic expansion of $F_1(x)$ given in 
\cite{Smirnov:1998vk}, as discussed before.

\section{Expansions for the Master Integrals $F_2$ and $F_3$}
\label{SecMI}

The second MI $F_2$ is directly determined from the first one by means of the 
(algebraic) equation (\ref{eqs}) with $i \, = \, 1$, which gives
\beq
F_2(x) \, = \, \frac{1}{2} \, x \, \frac{ d F_1 }{ d x } (x) \, + \, F_1(x),
\eeq
in which $F_1$ and $F_1'$ are assumed to be known.
One just substitutes the power-series expansions obtained so far for $F_1(x)$
and obtains series representations for $F_2(x)$ in the same convergence domains.

The third MI satisfies a first-order equation of the form
\beq
\frac{d F_3}{d x} (x) \, + \, \frac{1}{x} \, F_3(x) 
\, + \, N(x) \, = \, 0,
\eeq
where the known term can be written as:
\bea
N(x) &=& \frac{1}{6x}  F_1(x) 
 -  \frac{1}{3} \left( 1 + \frac{1}{x} \right)  F_2(x)  
 +  \frac{1}{96 x^2}  H(-1,0,0; x)
 +  \frac{1}{48 x^2}  H(r,r,0; x) \nn\\
&& 
 +  \frac{\pi^2}{192x^2 }  H(-1;  x). 
\eea
It involves harmonic polylogarithms as well as 
$F_1$ and $F_2$, which are assumed to be known\footnote{
It is interesting to note that formally $F_1$ and $F_2$
appear as sub-topologies in the evaluation of $F_3$.
That also implies that the general method of series expansions
allows to compute topologies with subtopologies themselves given by
series expansions.
}.
The associated homogeneous equation, 
\beq
\frac{ d  F_3^{(0)} }{d x} \, + \, \frac{1}{x} \, F_3^{(0)}(x) \, = \, 0,
\eeq
has the general solution
\beq
F_3^{(0)} (x) \, = \, \frac{K}{x},
\eeq
where $K$ is an integration constant.
A particular solution of the inhomogeneous first-order equation is found
in a straightforward manner with the general method of the variation 
of the constants:
\beq
\bar{F}_3 (x) \, = \, - \, \frac{1}{x} \, \int_0^x x' N(x') dx'.
\eeq
Since $F_1$ and $F_2$ are not given in closed form in terms of known
transcendental functions, but only in terms of (truncated) series
expansions, one has also to expand all the GHPLs appearing in $N(x)$.
As in the case of the first MI $F_1$, the general solution is: 
\beq
F_3 (x) \, = \, \frac{K}{x} + \bar{F}_3 (x).
\eeq
Analogously to the case of $F_1$, one can determine analytically
all the coefficients of the expansion of $F_2$ around $x = 0$,
by fixing
\beq
K \, = \, 0
\eeq
in order to avoid infrared power divergencies.
The matching procedure is similar to the one used for $F_1$; the main 
difference is that in this case there is only one undetermined coefficient
for each series because the differential equation is of first order.
Details can be found in \cite{tesigrassi}.

\section{Relation to the Equal-Mass Sunrise}
\label{relsun}

There is a conspicuous relation between
the homogeneous differential equation for the first master integral $F_1$
and the homogeneous equation for the equal mass sunrise in
two dimensions,
\beq
S(z) \, = \, \int 
\frac{1}{D_1  D_2  D_3}
\mathcal{D}^2 k_1  \mathcal{D}^2 k_2,
\eeq
where:
\bea
D_1 &=& k_1^2  +  m^2,
\\
D_2 &=& k_2^2  +  m^2,
\\
D_3 &=& (p  -  k_1  -  k_2)^2  +  m^2 ,
\eea
and
\beq
z =  - \frac{s}{m^2},
\eeq
with $s  = \, - \, p^2$ and $p^{\mu}$ the external momentum.
The homogeneous equation (see eq.~(4.3) of \cite{Laporta:2004rb}) reads:
\beq
\frac{ d^2 }{ d z^2 }  S(z;  0)
 + \left(
\frac{1}{z}  +  \frac{1}{z+1}  +  \frac{1}{z+9}
\right)
 \frac{ d }{ d z }  S(z;  0)
 + \left[
\frac{1}{3z}
 -  \frac{1}{4(z+1)}  -  \frac{1}{12(z+9)} 
\right] \, S(z)
\, = \, 0.
\eeq
By writing
\beq
F_1(x) = \frac{1}{x} \, M(-x-1),
\eeq
whose inverse reads:
\beq
M(z) = - (z+1) \, F_1(- z - 1),
\eeq 
we find that $M(z)$ satisfies the same homogeneous equation of $S(z)$.
The ``physical origin'' of such relation is unclear to us.

\section{Conclusions}
\label{concl}

We have computed the three master integrals for the crossed
ladder diagram with two equal-mass quanta exchanged by
means of various power-series expansions centered around
different points.
The two-equal-mass exchanged non-planar topology is by far the most complicated 
one entering the two-loop form factor and a novel structure emerges, related to the
occurrence of elliptic integrals rather than (generalized) harmonic polilogarithms.
That was not expected {\it a priori}, because the diagram
(see fig.~1) has only thresholds in $s \, = \, 0$ and in $s \, = \, m^2$,
as well as a pseudothreshold in $s = - 4 m^2$; we may only generically relate such 
structure to the non-planar topology.
We argue that the occurrence of elliptic integrals is probably a
general feature of massive multi-loop diagrams beyond some level 
of complexity.

By studying the diagram close to the massless threshold $s \, = \, 0$ we have
obtained a small momentum expansion with coefficients analitycally determined.
By combining our large-momentum expansion with that obtained in ref.\cite{Smirnov:1998vk},
we have been able to give the leading power-suppressed corrections to the logarithmic
contributions in a completely analytical form.

We have found an {\it a priori} unexpected relation between the basic (first) 
master integral for two-mass crossed ladder in four space-time dimensions and the basic master
integral for the equal mass sunrise in two dimensions: the homogeneous 
differential equation is exactly the same, while the inhomogeneous terms are unrelated
(in the crossed ladder case, the latter ones being, of course, much more complicated).

This work terminates the computation of the master integrals entering the 
two-loop electroweak form factor.

\vspace*{6mm}

\noindent {\large{\bf Acknowledgments}}

\vspace*{2mm}

\noindent 
R.~B. wishes to thank the Department of Physics of the University of 
Florence and INFN Section of Florence for kind hospitality during a part 
of this work. The work of R.~B. was partially supported by Ministerio de
Educaci\'on y Ciencia (MEC) under grant FPA2004-00996, 
Generalitat Valenciana under grant GV05-015, and MEC-INFN agreement.

\appendix

\section{Generalized Harmonic Polylogarithms}
\label{appa}

The general theory of the harmonic polylogarithms has already been discussed in
\cite{Remiddi:1999ew, Aglietti:2004tq, Aglietti:2004ki}, to which we refer for details.
Here we only give the specific information relevant to the computation
of the crossed ladder.
Let us first give the expressions of the harmonic polylogarithms (HPL's) entering our computation 
in terms of the standard ones:
\bea
H(-1; \, x) &=& \log (1 + x ) ;
\\
H(0,0;\, x) &=& \frac{1}{2} \, \log^2 (x) ;
\\
H(0,-1;\, x) &=& - \, {\rm Li}_2(-x) ;
\\
H(0,-1,0;\, x) &=& 2 {\rm Li}_3(-x) \, - \, \log x \, {\rm Li}_2(- x).
\eea
Let us now show that the Generalized Harmonic Polylogarithm (GHPL)\footnote{
The weight ``$r$'' labels the integration over the function
\beq
f(r;t) = \frac{1}{\sqrt{t(4-t)}} ,
\eeq
such that, for instance:
\beq
H(r;x) = \int_{0}^{x} \frac{dt}{\sqrt{t(4-t)}} \,.
\eeq
}
\beq
H(r,0; \, x) \, \equiv \, \int_0^{x} 
\frac{\log y }
{\sqrt{ y ( 4 \, - \, y )}} \, dy,
\eeq
is real for $0 < x < 4$, purely imaginary for $x>4$,  
while it is complex for $x<0$.
In fact, for $0 < x < 4$
the integrand is real in all the integration domain
and therefore also $H$ is real.
For $x>4$, since $x \to x-i \epsilon$ with $\epsilon=+0$ and, therefore,
$\sqrt{4-x+i \epsilon} \to i \sqrt{x-4}$,
one can split the integral as:
\beq
H(r,0; \, x) \, = \, \int_0^4 \frac{\log y}{\sqrt{y ( 4 \, - \, y)}} \, dy
\, - \, i \, \int_4^x \frac{\log y}{\sqrt{y ( y \, - \, 4)}} \, dy.
\eeq
The first integral on the r.h.s. vanishes,
\beq
\int_0^4 \frac{\log y}{\sqrt{ y ( 4 \, - \, y ) } } \,dy \,  = \, 0,
\eeq
implying that $H$ is purely imaginary in this region.
For $x < 0$ the integrand is complex and so is the $H$.
As a consequence, the term entering $C(x)$, the inhomogeneous 
(and known) term in the differential equation for $F_1$ given in eq.(\ref{eq54}),
\beq
\label{nosing}
\frac{ H(r,0; \, x) }{ \sqrt{x ( 4 \, - \, x ) } }
\eeq
is real for $x>0$, as it should, while it is complex
for $x<0$.

Let us now show that $H(r,r,0; \, x)$ is real for
$x>0$ while it is complex for $x<0$. Let us write:
\beq
H(r,r,0; \, x) \, \equiv \, 
\int_0^x \frac{1}{\sqrt{ y ( 4 \, - \, y ) } } 
\, H(r,0; \, y) \, dy.
\eeq
For $0<x<4$, the integrand is real in all the integration domain,
while for $x>4$ one can write:
\beq
H(r,r,0; \, x) \, = \, \int_0^4 \frac{1}{\sqrt{ y ( 4 \, - \, y ) } } H(r,0; \, y) \, dy
\, - \, i \, \int_4^x \frac{1}{\sqrt{ y ( y  \, - \, 4 ) } } H(r,0; \, y) \, dy.
\eeq
Both integrals on the r.h.s. are real.

\noindent
$H(r,0;\, x)$ can be expressed in terms of ordinary harmonic polylogarithms
of a non linear function of $x$, by changing the integration variable from
$y$ to
\beq
t \, =  \, \frac{y}{2} \, - \, 1 \, + \, \sqrt{y \, \left( \frac{y}{4} \, - \, 1 \right)}
\, = \, y \, - \, 2 \, - \, \frac{1}{y} \, - \, \frac{2}{y^2} \, - \, \frac{5}{y^3}
\, + \, O\left( \frac{1}{y^4} \right),   
\eeq
where the square root is the arithmetical one for $y \, > \, 4$, and on the last member
we have made the expansion for $y \, \gg \, 1$. The inverse reads:
\beq
y \,  = \, \frac{(t \, + \, 1)^2}{t}.
\eeq
Note that $y = 4 \, \leftrightarrow t = 1$. 
We obtain:
\beq
H(r,0; x)  =  
-  2  i  H\left[ 0, - 1;  \frac{x}{2} \! - \! 1 \! + \! 
\sqrt{x\left(\frac{x}{4} \! - \! 1\right)} \, \right]
 +  i  H\left[0,0;  \frac{x}{2} \! -\!  1 \! + \! 
 \sqrt{x\left(\frac{x}{4} \! - \! 1\right)} \, \right] 
 + \! i  \frac{\pi^2}{6}.
\eeq
In a similar way, we obtain for the only GHPL of weight 3 the representation:
\bea
H(r,0,0; x) \! \! & = & \! \! 
-  H\left[ 0, 0, 0 ;  \frac{x}{2} \! - \! 1 \! + \! \sqrt{x\left(\frac{x}{4} \! - \! 1\right)} \, \right]
 +  4  H\left[0,-1,0;  \frac{x}{2} \! - \! 1 \! + \! \sqrt{x\left(\frac{x}{4} \! - \! 1 \right) } \, \right] \, +
\nn\\
& &  + \, 6 \, \zeta(3),
\eea
where $H(0,-1,0;1) \, = \, - \, 3/2 \, \zeta(3)$.  
In a completely analogous manner one can analyze $H(r,r,0;x)$,
the GHPL entering the know term of the differential equation for $F_3$.

\section{Expansion of Generalized Harmonic Polylogarithms}
\label{appb}

The points $x \, = \, 0, \, 4$ and $\infty$ are singular points for $H(r,0; \, x)$.
That is easily seen just by looking at the integral definition.
The Taylor expansion of $H(r,0;\, x)$ centered around a {\it regular} point $x_0 \ne 0, \, 4, \, \infty$ 
is easily obtained by remembering that 
\beq
\frac{d}{d x} H(r,0;\, x) \, = \, f(r,0;\, x),
\eeq
where
\beq
f(r,0;\, x) \, = \, \frac{\log x}{\sqrt{x ( 4 \, - \, x ) }}.
\eeq
By performing the Taylor expansion of $f(r,0;\, x)$ and integrating on both
sides, one obtains:
\beq
H(r,0;\, x) \, = \, H(r,0;\, x_0)
\, + \, \sum_{n=0}^{\infty} \frac{1}{(n+1)!} \, f^{(n)}\left( r,0;\, x_0 \right) \, 
\left( x\, - \, x_0 \right)^{n+1}. 
\eeq
Let us note that one transcendental constant only is involved in this
expansion, namely $H(r,0;\, x_0)$.
Let us now consider the expansions around singular points.
It is clear that $H(r,0; \, x)$ does not possess an ordinary Taylor expansion around $x = 0$ --- 
the origin is a branch point of infinite order ---
where derivatives are not defined.
This GHPL however has an expansion around zero involving semi-integer powers 
of $x$ multiplied with up to one power of $\log x$, 
\beq
H(r,0;\, x) \, =  \, x^{1/2} \, \sum_{n=0}^{\infty} k_n \, x^n
\, + \, x^{1/2} \, \log x \, \sum_{n=0}^{\infty} l_n \, x^n,
\eeq
where $k_n$ and $l_n$ are coefficients and the serieses on the r.h.s. have
a convergence radius $R = 4$ --- i.e. up to the closest singularity.
The representation above can be obtained by expanding around $y=0$ the 
regular part of the integrand, i.e. the factor $ 1/\sqrt{4-y} $, so that
\beq
\frac{\log y}{\sqrt{y(4 \, - \, y)}} \, = \, \frac{1}{2} \, \log y \, y^{-1/2} 
\, + \, \frac{1}{16} \, \log y \, y^{1/2} 
\, + \, \frac{3}{256} \, \log y \, y^{3/2}
\, + \, \frac{5}{2048} \, \log y \, y^{5/2}
\, + \, O\left(y^{7/2}\right)
\eeq
and using the result:
\beq
\int \log y \, y^{\alpha} \, dy \, = \, 
\frac{ 1 }{ \alpha \, + \, 1 } \, \log y \, y^{\alpha \, + \, 1}
\, - \, \frac{ 1 }{ (\alpha \, + \, 1)^2 } \, y^{\alpha \, + \, 1}.
\eeq
The expansion around $x \, = \, 4$ is obtained in an analogous way, i.e.
by expanding the regular part of the integrand around $x = 4$, $\log y/\sqrt{y}$, and integrating
term by term. An expression of the following form is obtained:
\beq
H(r,0;x) \, = \, (x-4)^{1/2} \, \sum_{n=0}^{\infty} g_{n} \, (x-4)^{n},
\eeq
where 
\bea
g_{0} &=& 2 \, \log 2; \\
g_{1} &=& \frac{\log 2}{12}; \\
g_{2} &=& \frac{1}{80} \, - \, \frac{3}{320} \, \log 2 .
\eea
$x \, = \, 4$ is therefore a branch point of second order.
Let us note therefore that the expression (\ref{nosing})
has no singularities for $x\rightarrow 4$.

\noindent
By using the representation obtained in the previous appendix, we can directly obtain 
the expansion of the GHPL around infinity, which turns out to be of the form:
\beq
H(r,0;\, x) \, = \,
- \, i \, \sum_{n=0}^{\infty} \frac{a_n}{x^n}
\, - \, i \, \log x \, \sum_{n=0}^{\infty} \frac{b_n}{x^n}
\, - \, i \, \frac{1}{2} \, \log^2 x,
\eeq
where the two serieses on the r.h.s. are convergent for $|x| > 4$
and the lowest-order coefficients read:
\begin{align}
a_0 &= \frac{\pi^2}{6} ; &
b_0 &=  0 ;
\\
a_1 &= - \, 2 ;
&
b_1 &= - \, 2 ;
\\
a_2 &= - \, \frac{3}{2} ;
&
b_2 &= - \, 3 ;
\\
a_3 &= - \, \frac{20}{9} ;
&
b_3 &= - \, \frac{20}{3} .
\end{align}

\end{document}